\documentclass[largeformat]{interact} 
\usepackage{color,ulem,url}

\begin{document}
\newcommand\be{\begin{equation}}
\newcommand\ee{\end{equation}}
\newcommand\bea{\begin{eqnarray}}
\newcommand\eea{\end{eqnarray}}



\title{Spontaneous synchronization and nonequilibrium statistical mechanics of
coupled phase oscillators}

\author{
\name{Stefano Gherardini\textsuperscript{a},
Shamik Gupta\textsuperscript{b} and Stefano
Ruffo\textsuperscript{c}$^{\ast}$\thanks{$^\ast$Corresponding author; e-mail: ruffo@sissa.it}}
\affil{\textsuperscript{a}Department of Physics, University of Florence, via S. Marta 3, I-50139 Florence, INFN and LENS, via G. Sansone 1, I-50019 Sesto Fiorentino, Italy \\ 
\textsuperscript{b}\mbox{Department of Physics, Ramakrishna
Mission Vivekananda University, Belur Math, Howrah 711202, India} \\
\textsuperscript{c}SISSA, via Bonomea 265, CNISM and INFN, I-34136 Trieste, Italy}
}
\maketitle

\begin{abstract}
Spontaneous synchronization is a remarkable collective effect observed
in nature, whereby a population of oscillating units, which have diverse
natural frequencies and are in weak
interaction with one another, evolves to spontaneously exhibit
collective oscillations at a common frequency. The Kuramoto model
provides the basic analytical framework to study spontaneous
synchronization. The model comprises limit-cycle oscillators with distributed natural frequencies
interacting through a mean-field coupling. Although more than forty
years have passed since its
introduction, the model continues to occupy the
centre-stage of research in the field of non-linear dynamics, and is also widely applied to model diverse physical situations. In this
brief review, starting with a derivation of the Kuramoto model and the
synchronization phenomenon it exhibits, we summarize recent results on the study of
a generalized Kuramoto model that includes inertial effects and
stochastic noise. We describe the dynamics of the generalized model from a
different yet a rather useful perspective, namely, that of long-range interacting systems driven out of equilibrium by
quenched disordered external torques. A system is said to be long-range
interacting if the inter-particle potential decays slowly as a function
of distance. Using tools of statistical physics, we highlight the
equilibrium and nonequilibrium aspects of the dynamics of the
generalized Kuramoto model, and uncover a rather rich and complex
phase diagram that it exhibits, which underlines the basic theme of intriguing emergent
phenomena that are exhibited by many-body complex systems.
\end{abstract}

\tableofcontents

\section{Introduction: Spontaneous synchronization}
\label{sec:introduction}
Spontaneous synchronization is a general phenomenon in which a population of coupled oscillators
(usually of different frequencies) self-organizes to operate in unison~\cite{Pikovsky:2001,Strogatz:2003,Rosenblum:2003,Pikovsky:2007}. The
phenomenon is observed in physical and
biological systems over a wide range of spatial and temporal scales,
e.g., metabolic synchrony in yeast cell suspensions~\cite{Bier:2000},
flashing fireflies~\cite{Buck:1988}, Josephson junction
arrays~\cite{Wiesenfeld:1998}, laser arrays~\cite{Hirosawa:2013}, and others. Besides the synchronous firings of cardiac cells
that keep the heart beating and life going on~\cite{Winfree:1980}, synchrony is desired in many man-made systems, e.g., in
parallel computing, whereby computer processors must coordinate to finish a task on time, and in electrical
power-grids, in which generators must run in synchrony to be locked to
the grid frequency~\cite{Filatrella:2008,Rohden:2012}. Synchrony could
also be hazardous, e.g., in neurons, leading to impaired brain function
in Parkinson's disease and epilepsy. Collective synchrony in oscillator networks has attracted
immensely the attention of physicists and applied mathematicians, and finds applications in many fields, from
quantum electronics to electrochemistry, from bridge engineering to
social science. 

This paper provides a basic overview of the field of synchronization
from the point of view of a paradigmatic model for analytical studies,
the Kuramoto model. The model comprises limit-cycle oscillators with distributed natural frequencies
interacting through a mean-field coupling~\cite{Kuramoto:1975}. Since its
introduction about forty years ago, the model has been widely employed in the arena of non-linear dynamical system
studies to study the phenomenon of spontaneous synchronization, and continues to
inspire new expedition to the kingdom of many-body complex
systems. This brief review starts with a summary of useful dynamical
features of synchronizing systems, followed by a discussion of how they
may lead to a derivation of the Kuramoto model. A detailed discussion
follows of the synchronization phenomenon exhibited by the model and
also its 
extended version in which the dynamics proceeds in
presence of stochastic noise. We devote the rest of the paper to a study of
a generalized Kuramoto model that includes inertial effects and
stochastic noise, thereby elevating the first-order dynamics of the Kuramoto model to one that is second order in time. We describe the dynamics of the generalized model from a
different yet a rather useful perspective, namely, that of long-range interacting systems driven out of equilibrium by
quenched disordered external torques. This connection helps to study the
model from the point of view of statistical physics, besides offering to
form a bridge with a related but until now a largely unconnected
field of long-range interacting systems. In fact, we show that in proper
limits, the
generalized model quite remarkably reduces to the Kuramoto model as well
as to a prototypical system with long-range interactions, the
Hamiltonian mean-field model~\cite{Ruffo:1995}. Using tools of statistical physics, we highlight the
equilibrium and nonequilibrium aspects of the dynamics of the
generalized Kuramoto model. Further, we uncover a rather rich and complex
phase diagram that the model exhibits, demarcating regions of parameter space
that allow for the emergence of spontaneous synchronization.

The paper is organized as follows. In
Section~\ref{sec:theoretical-modelling}, we discuss some general
features of synchronizing systems and the derivation of
the Kuramoto model. In
Section~\ref{sec:synchronization-kuramoto}, we discuss the
analysis of the model in the thermodynamic limit, thereby obtaining the
conditions on the parameters of its dynamics that allow for the
observation of collective synchrony and an associated phase transition
in the stationary state. Here, we also discuss the
case of the noisy Kuramoto model. Section~\ref{sect:kuramoto-first-order}
contains detailed discussions on the
generalized Kuramoto model. The paper ends with conclusions in
Section~\ref{sec:conclusions}. 

\section{Theoretical modelling: From limit cycles to the Kuramoto
model}
\label{sec:theoretical-modelling}
It is clear from the aforementioned examples of synchronizing systems
that their constituent units are capable of exhibiting oscillations that have a characteristic
waveform, amplitude and frequency of oscillation. The latter features depend of course on the physical manifestation of the unit:
the heart does not beat the same way as a firefly flashes on
and off. Moreover, these characteristic oscillations are such that any (slight)
perturbations away from them would soon return the motion to the oscillatory behavior.  The
dynamics of the individual units should therefore be such as to allow
for oscillations that have a characteristic waveform independent of any
typical initial condition of the dynamics. Think of the pendulum of a metronome: irrespective of the initial deflection of its pendulum
(provided it is not so drastic that you break the metronome !), the
latter would soon tick and tock back and forth at a
given period, exhibiting oscillations that have both a characteristic amplitude and a
characteristic frequency.

Now, one may wonder: How should the underlying dynamics be such as to generate oscillations with a characteristic waveform? On the
basis of physical intuition, one may
anticipate (correctly) that the dynamics ought to have 
suitable dissipation and energy-pumping mechanisms so that oscillations that tend to
become too large are effectively damped down by dissipation, just as the ones that tend
to become too small are suitably pumped up by a supply of energy. As a
result, oscillations of a characteristic form, for which pumping and damping effects balance each
other, are only
sustained. The presence of damping at once precludes the
possibility for the underlying dynamics to be conservative, i.e., a
dynamics given by the Hamilton equations of motion corresponding to a
suitable system Hamiltonian. Consequently, the stationary state that the
dynamics relaxes to at long times would not be an equilibrium
one, but would be a generic nonequilibrium stationary state (NESS)~\cite{Livi:2017}. The
reader may recall that the basic tenet of classical equilibrium statistical
mechanics is a dynamics modelled by the Hamilton equations of motion derived
from the Hamiltonian of the system under consideration.

\begin{figure}[ht]
 \centering
 \includegraphics[width=\textwidth]{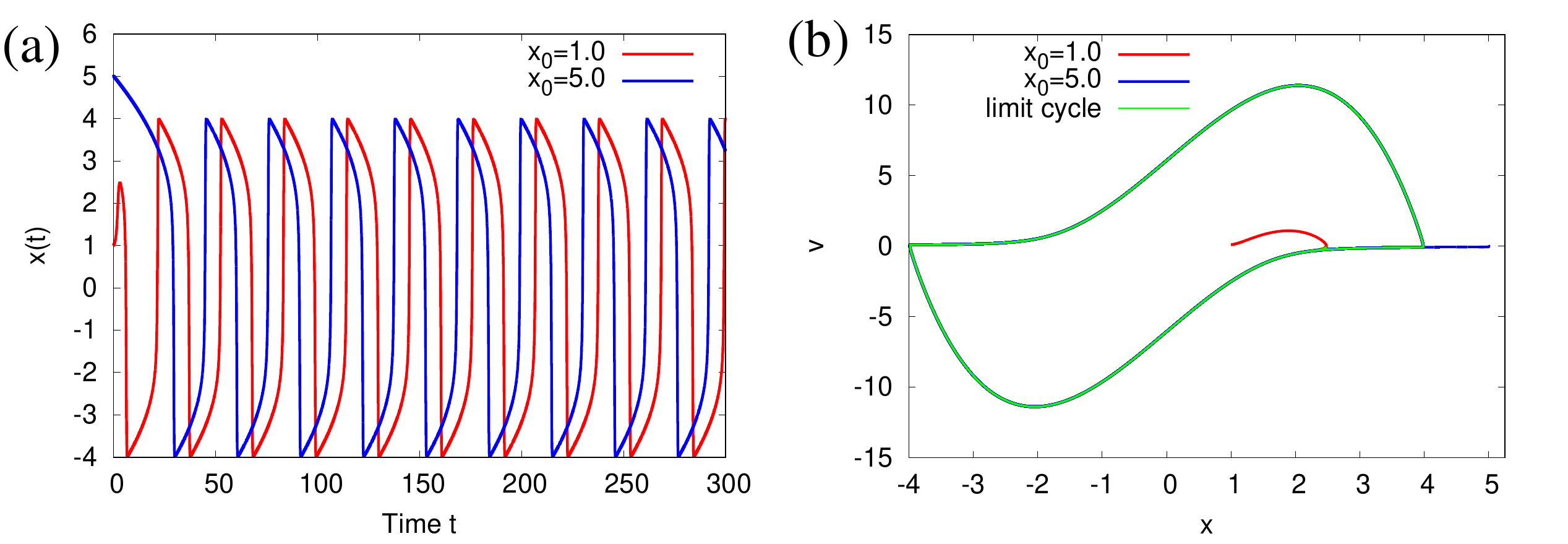}
 \caption{The Van der Pol oscillator dynamics, Eq.~(\ref{eq:damped-driven-oscillator}) (equivalently, Eq.~(\ref{eq:damped-driven-oscillator-xv})), showing for two different
 initial values $x_0$ of the position and a given initial value
 $v_0=0.1$ of the
 velocity (a) the displacement
 $x$ as a function of time, and (b) the trajectory traced out in the
 phase space $(x,v)$ by the initial point $(x_0,v_0)$. In panel (a), one may
 observe that independently of initial conditions, the dynamics in a short time settles into oscillations
 of a characteristic waveform. Correspondingly, one has in panel (b) an
 eventual relaxation to a motion along a limit cycle; for large times,
 the motion is virtually indistinguishable from the limit cycle. The
 data are obtained by numerical integration of Eq.~(\ref{eq:damped-driven-oscillator-xv}) for parameter values $\gamma=4.0,\omega=0.5$.}
 \label{fig:vanderpol}
\end{figure}

Let us illustrate with an example how a dynamics that
incorporates dissipation and energy-pumping mechanisms leads to
oscillations of a characteristic form independent of initial conditions. Consider a single dynamical
degree of freedom $x(t)$ describing the displacement from equilibrium of
a damped, driven harmonic
oscillator, whose time evolution is given by the so-called Van der Pol equation:
\begin{equation}
\label{eq:damped-driven-oscillator}
\frac{{\rm d}^2 x}{{\rm d}t^2}-(\gamma-x^2)\frac{{\rm d}x}{{\rm d}t}+\omega^2x=0\,.
\end{equation}
Here, the parameter $\gamma$ is a real positive constant, while the
parameter $\omega$ is real. In the dynamics~(\ref{eq:damped-driven-oscillator}), note that the second term changes sign depending on whether $x$ has a magnitude
smaller or larger than a characteristic value equal to $\sqrt{\gamma}$.
As a result, the dynamics pumps up small displacements (i.e., with
$|x| < \sqrt{\gamma})$ and damps down the large ones (i.e., with $|x| >
\sqrt{\gamma}$). Hence, independently of initial conditions, the
dynamics for given values of $\gamma$ and $\omega$ approaches
asymptotically in time a state that supports oscillations
with a characteristic amplitude and a characteristic frequency. In other
words, the solution $x(t)$ of the dynamics~(\ref{eq:damped-driven-oscillator}) becomes in the long-time limit a
periodic motion with a characteristic waveform, see Fig.~\ref{fig:vanderpol}(a).

\begin{figure}[ht]
 \centering
 \includegraphics[width=\textwidth]{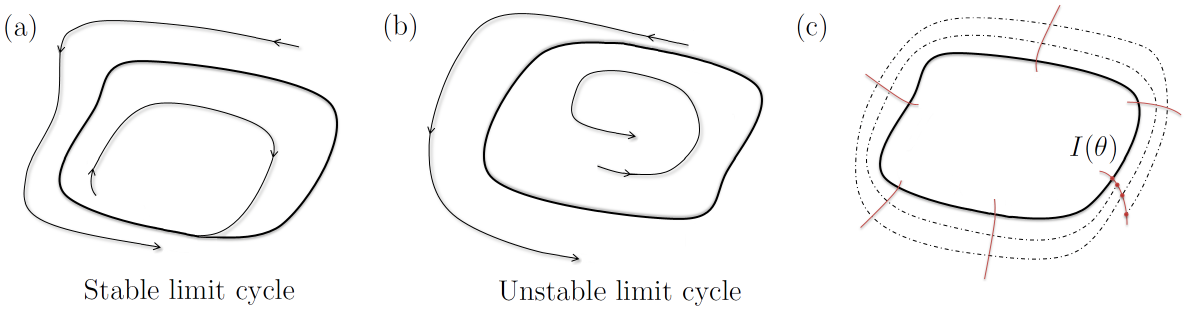}
 \caption{Panels (a) and (b) compare the dynamics around a stable and an
 unstable limit cycle. While nearby trajectories under dynamical
 evolution are attracted towards a stable limit cycle, they are
 instead repelled away from the cycle when it is unstable. In panel
 (c), we illustrate the construction of isochrones for a stable limit
 cycle.}
 \label{fig:limitcycle}
\end{figure}

The dynamics~(\ref{eq:damped-driven-oscillator}) may be written
in terms of a set of two coupled first-order differential equations, by
introducing the velocity variable $v$ as follows:
\begin{equation}
\frac{{\rm d}x}{{\rm d}t}=v\,,~~
\frac{{\rm d}v}{{\rm d}t}=(\gamma-x^2)v-\omega^2 x\,. 
\label{eq:damped-driven-oscillator-xv}
\end{equation}
Then, in the phase space of the system, given by the two-dimensional
plane $(x_1,x_2)\equiv(x,v)$, the dynamical trajectory/orbit traced out by an
initial point $(x_0,v_0)$ may be seen to approach asymptotically in time a stable periodic
orbit that is in one-to-one mapping with the long-time periodic solution
discussed above. Oscillations that have a characteristic waveform, amplitude
and frequency, and are thus represented by a characteristic periodic
orbit in the phase space are said to define the so-called limit-cycle
oscillators~\cite{Strogatz:2014}. We may thus say that any initial condition evolving under
the dynamics~(\ref{eq:damped-driven-oscillator}) eventually
relaxes to a motion around a limit cycle given by the aforementioned
periodic orbit. In particular, an orbit starting close to
the limit cycle gets after a very short time extremely
close to the cycle and becomes essentially indistinguishable from the latter,
although mathematically speaking, it never reaches it due to
the uniqueness of solutions of the dynamics. The limit cycle is stable in the sense that any (small)
perturbations away from it decay in time, thereby attracting all
neighboring orbits towards it under the dynamical evolution. A
limit cycle can also be unstable, whereby all neighboring orbits
are repelled away from it under dynamical evolution~\cite{Strogatz:2014}.
Figure~\ref{fig:limitcycle}, panels (a) and (b) compare a stable and an unstable limit
cycle. The limit cycle for dynamics~(\ref{eq:damped-driven-oscillator}) is
shown in Fig.~\ref{fig:vanderpol}(b).

A limit cycle is evidently an ${\it
isolated}$ periodic orbit in the phase space. These cycles can
occur only in non-linear dynamical systems. A linear dynamics ${\rm
d}x_\alpha/{\rm d}t=\sum_{\alpha,\beta}A_{\alpha \beta}x_\beta$ can of
course generate periodic orbits, but since with every
periodic orbit $\{x_\alpha(t)\}$, one may associate a
family of periodic orbits $\{cx_\alpha(t)\}$ with $c$ a parameter, such an orbit would not be isolated, but
would be surrounded by an infinite number of periodic orbits obtained by
varying $c$. The
issue of which one among the orbits is chosen by the dynamics is set by its
initial condition, unlike the independence of the form of
a limit cycle with respect to initial conditions. Also, any slight
perturbation away from such a closed orbit will unlike a limit
cycle not return the motion to the orbit, but will take it to
a neighboring closed orbit.

The above comments on the definition and properties of a limit cycle
and the nature of the dynamics leading to it also apply to
a generic autonomous dynamical system comprising many interacting
degrees of freedom $\{x_\alpha\}_{1\le \alpha \le n};~n \gg 1$, with a
dynamics given by~\cite{Strogatz:2014}
\begin{equation}
\frac{{\rm d}x_\alpha}{{\rm d}t}=F_\alpha(x_1,x_2,\ldots,x_n)\,.
\label{eq:generic-dynamical-system}
\end{equation}
By autonomous is meant that the functions $F_\alpha$ do
not depend explicitly on time. 

For a given initial condition $\{x_\alpha(0)\}$, a solution $\{x_\alpha(t)\}$
of the dynamics~(\ref{eq:generic-dynamical-system}) defines an orbit in the
$n$-dimensional phase space of the system. Being an autonomous dynamics
implies that if $\{x_\alpha(t)\}$ is a solution, so is $\{x_\alpha(t+t_0)\}$ for
any $t_0$, i.e., the choice of the origin of
time is irrelevant. Such a property holds either when there
is no external influence on the system so that its 
motion depends solely on the interaction between its constituents, or, even when there is an external
influence, it does not depend explicitly on time.
Limit cycles denote a
particular class of solutions $\{x_{\alpha,0}(t)\}$ represented by one-dimensional periodic orbits in
the phase space that satisfy
$x_{\alpha,0}(t+T)=x_{\alpha,0}(t)~\forall~\alpha$, where $T$ is the period
of the motion. In the following, any mention of oscillator would
mean a stable limit-cycle oscillator, unless stated otherwise.

Consider a limit cycle with period $T$. The length of the orbit 
traversed in the phase space in time $t$ is given by $\displaystyle s(t)=\int_0^t {\rm d} t' \sqrt{ \sum_{\alpha=1}^n
\left({\rm d}x_\alpha/{\rm d}t\right)^2(t')}$, where the freedom in choosing the origin of time
translates to the one in the choice of the origin from which
the orbit length is measured. Note that $s(pT)=ps(T)$, where $p$
is a positive integer, and that the time rate of variation of
$s$ is in general not a constant along the limit cycle. One may however transform to a new variable $\theta=\theta(s)$ whose
time rate of variation along the limit cycle is a constant called the
{\it natural frequency} $\omega \equiv 2\pi/T$ of the cycle. Using
$\displaystyle \theta(s)\equiv (2\pi/T)\int_0^s~{\rm d} s'/\left[\left(\frac{{\rm
d}s}{{\rm d}t}\right)(s')\right]$, we have indeed ${\rm d}
\theta/{\rm d} t=({\rm d} \theta/{\rm d}s)({\rm d} s/{\rm d} t)=2\pi/T=\omega$. Moreover, defining $s_0\equiv s(T)$,
one has $\theta(s_0)=2\pi$. Thus, at the end of one time period $T$,
the value of $\theta$ increases by $2\pi$, corresponding to one complete
traversal of the periodic orbit. We thus arrive at an important conclusion that a limit-cycle
oscillator is completely characterized by a phase $\theta \in [-\pi,\pi]$ that changes uniformly in time with period $T$ and frequency
$\omega$, as
\begin{equation}
\frac{{\rm d}\theta}{{\rm d}t}=\omega\,.
\label{eq:dtheta-dt}
\end{equation}
In this paper, the word `phase' would be used to also refer to a
thermodynamic phase of a macroscopic system, defined as a region in the
space of dynamical parameters throughout which all macroscopic
observable properties of the system are essentially
the same. To avoid any possible confusion between the two different usages
of the word `phase,' we will from now on use the term `angle' to mean oscillator phase, and
the term `phase' to exclusively mean a thermodynamic phase.

Since we may associate a unique value of the angle with each point
on the limit cycle, we have $\theta=\theta(\{x_{\alpha,0}\})$.
From Eq.~(\ref{eq:dtheta-dt}), it follows that $\theta$ is a neutrally
stable variable: any (small)
perturbations to
it neither grow nor decay in time. This property is related to the
invariance discussed above of solutions of autonomous dynamical
systems with respect to time shifts. In contrast to $\theta$, the amplitude of oscillations has a definite
stable value on the limit cycle; any (small) perturbations in a direction transverse to
the angle decay in time.

An angle description such as above applies even to 
orbits that are close to the limit cycle. To see this, consider an initial
phase-space point sufficiently close to the limit cycle. As the point
traverses an orbit in the phase space, we may decide to observe its
successive positions only stroboscopically, namely, at times $t=kT;
~k=1,2,3,\ldots$. It follows from the attracting property of the limit cycle that the limit of this sequence of points as $k \to \infty$ is a point on the
limit cycle, which according to the discussion above has a particular
value of the angle $\theta$. One may then associate this latter value
of $\theta$ with the sequence of points, which are now said to lie on a
$(n-1)$-dimensional hypersurface $I(\theta)$ called an isochrone~\cite{Pikovsky:2001}.
Figure~\ref{fig:limitcycle}(c) illustrates the construction of isochrones. In this way, we may
associate an angle $\theta$ to each point of the phase space lying in
close neighborhood of the limit cycle, and consequently, Eq.~(\ref{eq:dtheta-dt}) remains valid also in close neighborhood of the limit
cycle. 

Having explained the angle description in the context of individual limit-cycle oscillators, we now
make an observation that will prove to be quite relevant in our later
discussion on oscillators interacting
weakly with one another. To this end, consider a limit-cycle oscillator subject to
a forcing that is {\it weak}, which could be due to an
external agent or is generated due to the interaction of the oscillator with other
oscillators. Owing to the neutral
stability of the angle of the limit cycle, even a weak forcing can result
in it undergoing {\it large} changes. This may be contrasted with the
corresponding effect on the amplitude of oscillations of the limit cycle, which due to the transversal
stability of the cycle is only slightly affected by the external forcing. As a result, even
in the presence of a forcing, so long as it is weak, one is
justified to continue charactering the dynamics of the oscillator solely
in terms of an angle motion along the limit cycle of the isolated system, and to disregard to
leading order any perturbation due to the forcing in a
direction transversal to the isolated cycle. This so-called phase approximation for weak forcing allows to derive the
dynamics of a population of nearly-identical weakly-interacting limit-cycle
oscillators~\cite{Nakao:2015,Kuramoto:1984}, as we now do in the following.

Consider a collection of $N$ nearly-identical
limit-cycle oscillators occupying the nodes of a network and interacting
weakly with one another, with all the
oscillators having the same number $n$ of degrees of freedom. Since the oscillators are nearly identical, they will have dynamical properties that are only slightly different from
one another, with differences being of $O(\varepsilon)$, where
$\varepsilon$ is a small parameter. In the following, we use Greek letters to denote the different degrees
of freedom and Latin letters to denote the different oscillators. The
$j$-th oscillator, $j=1,2,\ldots,N$, having degrees of
freedom denoted by the set $\{x_{j\alpha}\}_{1 \le \alpha \le n}$, may
be 
considered to have the time evolution given by
\begin{equation}
\frac{{\rm d}x_{j\alpha}}{{\rm
d}t}=F_{j\alpha}(\{x_{j\beta}\})+\varepsilon\sum_{k=1,k\ne j}^N\sum_{\beta=1}^n
G_{j\alpha,k\beta}(\{x_{j\gamma}\},\{x_{k\gamma}\})\,,
\label{eq:xj-evolution}
\end{equation}
where the functions $F_{j\alpha}(\{x_{j\beta}\})$ describe the dynamics of
the isolated oscillators, while the function
$G_{j\alpha,k\beta}(\{x_{j\gamma}\},\{x_{k\gamma}\})$ represents the influence of
the $k$-th oscillator on the $j$-th one, with the small parameter
$\varepsilon$ ensuring that the oscillators are interacting only weakly
with one another. From Eq.~(\ref{eq:xj-evolution}), it is evident that
$1/\varepsilon$ has the dimension of time. Now, $\varepsilon$ being very
small, we may expect $1/\varepsilon$ to be longer than
any other characteristic timescale in the dynamics. 
Since the oscillators have dynamical properties that are only slightly
different, of $O(\varepsilon)$, we may write 
\begin{equation}
F_{j\alpha}(\{x_{j\beta}\})=F_\alpha(\{x_{j\beta}\})+\varepsilon
f_{j\alpha}(\{x_{j\beta}\})\,,
\end{equation}
expressing the heterogeneity of individual oscillators as small
fluctuations, denoted by $\varepsilon
f_{j\alpha}(\{x_{j\beta}\})$, about their common dynamical features
given by the functions $F_\alpha$. Equation~(\ref{eq:xj-evolution}) then yields
\begin{equation}
\frac{{\rm d}x_{j\alpha}}{{\rm
d}t}=F_\alpha(\{x_{j\beta}\})+\varepsilon \left[
f_{j\alpha}(\{x_{j\beta}\})+\sum_{k=1,k\ne j}^N \sum_{\beta=1}^n
G_{j\alpha,k\beta}(\{x_{j\gamma}\},\{x_{k\gamma}\})\right]\,.
\label{eq:xj-evolution-1}
\end{equation}

Now, let us assume that the common dynamics ${\rm d}x_{\alpha}/{\rm
d}t=F_\alpha(\{x_{\beta}\})$ allows for a stable limit cycle
characterized by the angle $\theta$, with the associated dynamical
degrees of freedom denoted by the set $\{x_{\alpha,0}\}$. The angle
$\theta$ evolves in time as ${\rm d}\theta/{\rm d}t=\omega$, where
$\omega$ is the natural frequency of the limit cycle of the
common dynamics; we denote the corresponding time period by $T\equiv 2\pi/\omega$. 

Consider a phase-space point $\{x_{j\beta}\}$ close to the limit
cycle of the common dynamics. As the phase-space point moves in time, it will owing to the smallness
of $\varepsilon$ continue to lie close to the limit cycle, moving between a family of isochrones $I(\theta_j)$ defined for the
limit cycle and characterized by different values of the angle. As a result, one has the functional
dependence $\theta_j=\theta_j(\{x_{j\beta}\})$. Using ${\rm
d}\theta_j/{\rm d}t=\sum_{\alpha=1}^n \left(\partial \theta_j/\partial
x_{j\alpha}\right)\left({\rm d}x_{j\alpha}/{\rm d}t\right)$ and
Eq.~(\ref{eq:xj-evolution-1}), we get 
\begin{equation}
\frac{{\rm d}\theta_j}{{\rm d}t}=\sum_{\alpha=1}^n
\left(\frac{\partial \theta_j}{\partial
x_{j\alpha}}\right)F_\alpha(\{x_{j\beta}\})+\varepsilon\sum_{\alpha=1}^n
\left(\frac{\partial \theta_j}{\partial
x_{j\alpha}}\right)\left[ f_{j\alpha}(\{x_{j\beta}\})+\sum_{k=1,k \ne j}^N \sum_{\beta=1}^n
G_{j\alpha,k\beta}(\{x_{j\gamma}\},\{x_{k\gamma}\})\right]\,.
\end{equation}
Comparing the form of the above equation with Eq.~(\ref{eq:dtheta-dt}), we find that the first term on the right
hand side (rhs) equals $\omega$, while to leading order in $\varepsilon$, one may replace the phase-space variables in
the second term with their values on the limit cycle. We thus obtain the
angle dynamics perturbed by the weak interaction among the oscillators
as
\begin{equation}
\frac{{\rm d}\theta_j}{{\rm d}t}
=\omega+\varepsilon
\sum_{\alpha=1}^nZ_{\alpha}(\theta_j)\Big[f_{j\alpha}(\theta_j)+\sum_{k=1,
k \ne j}^N \sum_{\beta=1}^n
G_{j\alpha,k\beta}(\theta_j,\theta_k)\Big];~~Z_{\alpha}(\theta_j)\equiv\left(\frac{\partial
\theta_j}{\partial
x_{j\alpha}}\right)(\{x_{\gamma,0}\})\,,
\label{eq:xj-evolution-2}
\end{equation}
where we have 
$f_{j\alpha}(\theta_j)=f(\{x_{\alpha,0}\}(\theta_j))$ and
$G_{j\alpha,k\beta}(\theta_j,\theta_k)=G_{j\alpha,k\beta}(\{x_{\alpha,0}\}(\theta_j),\{x_{\beta,0}\}(\theta_k))$.
The small-$\varepsilon$ approximation made in writing
Eq.~(\ref{eq:xj-evolution-2}) entails an error of order $\varepsilon^2$.

Let us introduce the difference between the
oscillator angles $\theta_j$ and the steadily increasing component
$\omega t$ corresponding to in-phase (synchronized) oscillations of all
the oscillators, as
\begin{equation}
\psi_j(t) \equiv \theta_j(t)-\omega t\,.
\label{eq:psij-definition}
\end{equation}
A time-independent $\psi_j$ implies that all the oscillators are oscillating in
synchrony with frequency $\omega$. In general, however, $\psi_j$ is time
dependent. Equations~(\ref{eq:psij-definition}) and
(\ref{eq:xj-evolution-2}) yield the time evolution of $\psi_j$ as
\begin{equation}
\frac{{\rm d}\psi_j}{{\rm d}t}=\varepsilon 
\sum_{\alpha=1}^nZ_{\alpha}(\psi_j + \omega t)\Big[f_{j\alpha}(\psi_j +
\omega t)+\sum_{k=1,k\ne j}^N \sum_{\beta=1}^n
G_{j\alpha,k\beta}(\psi_j+\omega t,\psi_k + \omega t)\Big]\,,
\label{eq:psij-evolution}
\end{equation}
which combined with the smallness of $\varepsilon$ implies that $\psi_j$
varies rather slowly with time, unlike the term $\omega t$ that varies
rapidly with time. In other words, suppose that at some instant, the
oscillators get synchronized with one another. Such a state will be sustained over times of order
$1/\varepsilon$ (which as mentioned above is the longest time interval in the system), during
which the term $\omega t$ would undergo a large number of changes, namely,
of order $\omega/\varepsilon$. As a result, over
the period $T = 2\pi/\omega$, one may consider all the $\psi_j$'s to be almost time independent, and so can average Eq.~(\ref{eq:psij-evolution}) over this period by considering the $\psi_j$'s
to be constant. We arrive at
\begin{eqnarray}
&&\frac{{\rm d}\psi_j}{{\rm
d}t}=\varepsilon\left[\Delta_j+\sum_{k=1,k \ne
j}^N\Gamma_{jk}(\psi_j-\psi_k)\right]\,;\label{eq:psij-eq}\\
&&\Delta_j \equiv \frac{1}{T}\int_0^{T} {\rm d}t'~\sum_{\alpha=1}^n
Z_\alpha(\psi_j+\omega t')f_{j\alpha}(\psi_j+\omega
t')\,,\\
&&\Gamma_{jk}(\psi_j-\psi_k)\equiv\frac{1}{T}\int_0^{T}{\rm
d}t'~\sum_{\alpha=1}^n \sum_{\beta=1}^n Z_\alpha(\psi_j+\omega
t')G_{j\alpha,k\beta}(\psi_j+\omega t',\psi_k + \omega t')\,.
\label{eq:Gammajk}
\end{eqnarray}
That the integral on the rhs of Eq.~(\ref{eq:Gammajk}) gives a function
of the angle difference may be inferred by noting that the angles
$\psi_j$ and $\psi_k$ are measured with respect to a zero-angle axis
that is arbitrary, and hence, one may choose to measure $\psi_k$ with
respect to $\psi_j$. In doing so, the rhs equals
$\displaystyle (1/T)\int_0^{T}{\rm
d}t'~\sum_{\alpha=1}^n \sum_{\beta=1}^n Z_\alpha(\omega
t')G_{j\alpha,k\beta}(\omega t',\psi_k-\psi_j + \omega t')$, which
evidently establishes the fact that the rhs of Eq.~(\ref{eq:Gammajk})
and hence, $\Gamma_{jk}$ is a function of the angle difference
$\psi_j-\psi_k$. Note that both the functions $Z(\psi)$ and
$\Gamma_{jk}(\psi)$ are $2\pi$-periodic in their argument.

Using Eqs.~(\ref{eq:psij-definition}) and~(\ref{eq:psij-eq}), we may
revert to the variables $\theta_j$, and obtain the corresponding dynamical
evolution as
\begin{equation}
\frac{{\rm d}\theta_j}{{\rm d}t}=\omega_j+\varepsilon
\sum_{k=1,k\ne j}^N\Gamma_{jk}(\theta_j-\theta_k)\,,
\label{eq:thetaj-evolution}
\end{equation}
where $\omega_j \equiv \omega+\varepsilon\Delta_j$ may be regarded as
the natural frequency of the $j$-th oscillator. The function
$\Gamma_{jk}(\theta)$, known as the phase coupling function, represents the effect of the $k$-th oscillator on the $j$-th one when
averaged over one period of limit-cycle oscillations of the common dynamics.

In the particular case when the function $G_{j\alpha,k\beta}$ is the same for all pairs
$(j,k)$ of oscillators, and has a magnitude of order $1/N$,
Eq.~(\ref{eq:thetaj-evolution}) reduces to the form
\begin{equation}
\frac{{\rm d}\theta_j}{{\rm d}t}=\omega_j+\frac{\varepsilon}{N}
\sum_{k=1, k \ne j}^N\Gamma(\theta_j-\theta_k)\,.
\label{eq:thetaj-evolution-all-to-all}
\end{equation}
The 
choice $\Gamma(\theta)=-{\cal K} \sin \theta$, with ${\cal K}$ being a
constant, reduces Eq.~(\ref{eq:thetaj-evolution-all-to-all}) to the dynamics of the
celebrated Kuramoto model of synchronization~\cite{Kuramoto:1975,Kuramoto:1984,Strogatz:2000,Acebron:2005,Gupta:2014-2,Gupta:2017-2}:
\begin{equation}
\frac{{\rm d}\theta_j}{{\rm d}t}=\omega_j-\frac{K}{N}
\sum_{k=1}^N\sin(\theta_j-\theta_k)\,,
\label{eq:kuramoto-model}
\end{equation}
where we have defined $K \equiv \varepsilon {\cal K}$. The sine function
in the last equation automatically
takes care of the fact that the summation on the rhs does
not include the term $k=j$. Moreover, the factor $1/N$ on the rhs
ensures that the net effect felt by one oscillator
due to all the other oscillators is independent of their total number,
thereby ensuring a well-defined behavior of the dynamics in the
thermodynamic limit $N \to \infty$. The constant $K$ characterizes the
strength of coupling between the oscillators. While in this work, we will discuss the version of the Kuramoto model as
in Eq.~(\ref{eq:kuramoto-model}) that involves time-independent
couplings, may we point out recent works of
interest, Refs.~\cite{Petkoski:2012,Pietras:2016}, on non-autonomous dynamics and time-varying
frequencies and couplings in the framework of the Kuramoto
model.

\section{Synchronization in the Kuramoto model and the associated phase transition}
\label{sec:synchronization-kuramoto}
The Kuramoto model is a dynamical system with $N$ interacting degrees
of freedom, and, as we will stress in the following, its invariant
measure in the thermodynamic limit $N \gg 1$ may be quite effectively
studied by using tools of statistical physics. In this
limit, let ${\cal G}(\omega)$ be the normalized number density of the
oscillator frequencies, i.e., the product ${\cal G}(\omega){\rm d}\omega$ gives the number
of oscillators whose natural frequencies lie in the range
$[\omega,\omega+{\rm d}\omega]$, with $\int{\rm d}\omega~{\cal
G}(\omega)=1$. In the language of statistical physics,
the $\omega_j$'s may be regarded as random variables sampled from the
underlying distribution ${\cal G}(\omega)$. Since the natural
frequencies for a set of oscillators have given values that are time
independent, $\omega_j$'s are to be regarded as {\it quenched
disordered} random variables, that is, those having values that do not evolve
in time. This may be contrasted with {\it annealed disorder} associated with
random variables whose values evolve in time.

The Kuramoto model has been mostly
studied for a unimodal ${\cal G}(\omega)$ with a non-compact support,
that is, one which is defined in the range $\omega \in [-\infty,\infty]$
and is symmetric
about the mean $\langle \omega \rangle\equiv \int_{-\infty}^\infty {\rm
d}\omega~\omega {\cal G}(\omega)$, and which decreases monotonically and
continuously to zero with increasing $|\omega-\langle \omega \rangle|$.
Of course, we assume here that ${\cal G}(\omega)$ is such that its mean
exists and is finite.
In this work, we will consider a ${\cal G}(\omega)$ that has all
the aforementioned properties.

\begin{figure}[ht]
 \centering
 \includegraphics[width=0.5\textwidth]{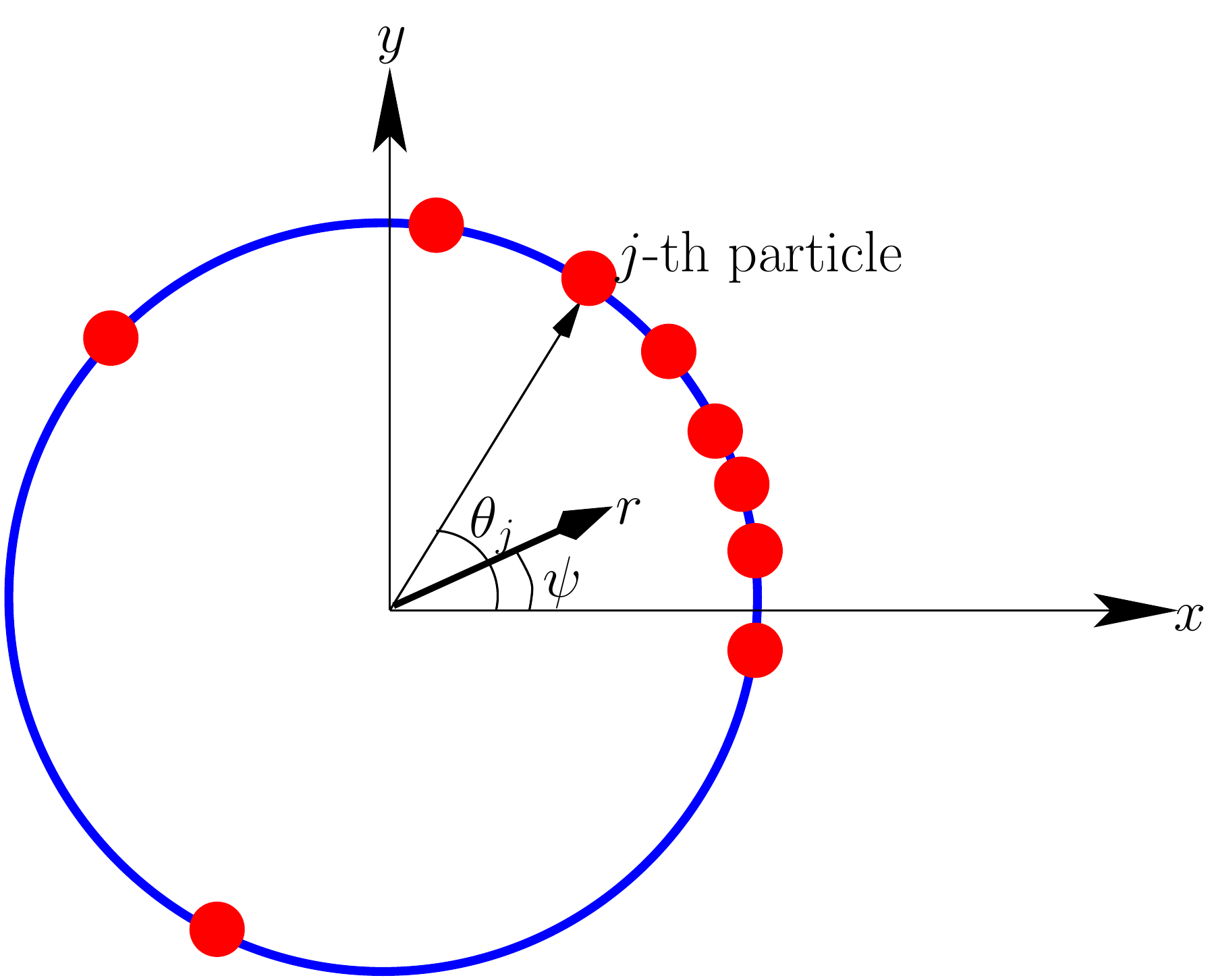}
 \caption{For the Kuramoto model of oscillators,
 Eq.~(\ref{eq:kuramoto-model}), the figure shows the
 quantities $r$ and $\psi$, see Eq.~(\ref{eq:rpsi-definition}), for a
 given configuration of oscillator angles $\theta_j$. As shown here, the centroid
 of the oscillator angles is given by the complex number $re^{i\psi}$.}
 \label{fig:rpsi}
\end{figure}

In the
dynamics~(\ref{eq:kuramoto-model}), although
a pair of oscillators is interacting rather weakly with one another due to the scaling of
the coupling $K$ by $N$, every oscillator is effectively responding to the {\it
collective} influence of all the other oscillators. To see this, it is
convenient to think of the oscillator angles as a collection of points moving on a unit
circle. Then, at any time $t$, one may associate a vector of unit length to each point, take a vector
sum, and divide by $N$, to get a vector of length $r(t)$ inclined at an
angle $\psi(t)$ with respect to a reference axis~\cite{Kuramoto:1984,Strogatz:2000}: 
\begin{equation}
r(t)e^{i\psi(t)}=\frac{1}{N}\sum_{j=1}^N e^{i\theta_j(t)}\,.
\label{eq:rpsi-definition}
\end{equation}
Here, $\psi(t)$ gives the
average angle, while $r(t)$ measures the amount of phase coherence or
synchrony in the system at time $t$, see Fig.~\ref{fig:rpsi}. Indeed, if
the angles are scattered around randomly on
the circle, one has $r(t)=0$, while, by contrast, if the oscillator angles are
clustered together on the circle, we have $r(t)>0$. In the
extreme case when all the oscillator angles have the same value, $r(t)$
attains its maximum possible value of unity. Referring to
Fig.~\ref{fig:rpsi}, we may express the quantity $r$ in terms of its
$x$ and $y$ components as
\begin{eqnarray}
&&r_x(t)=\frac{1}{N}\sum_{j=1}^N \cos(\theta_j(t))=r(t)\cos
\psi(t)\,,\nonumber \\
&&r_y(t)=\frac{1}{N}\sum_{j=1}^N \sin (\theta_j(t))=r(t)\sin
\psi(t)\,,\nonumber \\
&&r(t)=\sqrt{r_x^2(t)+r_y^2(t)},~~\psi(t)=\tan^{-1}(r_y(t)/r_x(t))\,.
\nonumber \\
\label{eq:rxry}
\end{eqnarray}
In terms of $r(t)$ and
$\psi(t)$, one may rewrite Eq.~(\ref{eq:kuramoto-model}) as
\begin{equation}
\frac{{\rm d}\theta_j}{{\rm d}t}=\omega_j-Kr(t)\sin(\theta_j-\psi(t))\,,
\label{eq:kuramoto-single}
\end{equation}
which puts in evidence the fact that every oscillator is being influenced by the
same combined effect expressed by
the quantities $r(t)$ and $\psi(t)$ generated due to all the
oscillators. Such a feature is generic to
statistical physical models with the so-called mean-field interaction in
which every constituent particle interacts
with all the other particles with the same strength. The
form~(\ref{eq:kuramoto-single}) makes this mean-field nature of the
dynamics evidently manifest. 

In passing, let us make a relevant observation. Let us consider Eq.~(\ref{eq:kuramoto-model}) and sum both sides over
$j$. We get
\begin{equation}
\frac{{\rm d}(1/N)\sum_{j=1}^N\theta_j}{{\rm d}t}=\frac{1}{N}\sum_{j=1}^N \omega_j\,,
\label{eq:kuramoto-model-sumj}
\end{equation}
which implies that considering the swarm of angle points moving
on the unit circle, their centroid turns around uniformly in time with a
frequency equal to $(1/N)\sum_{j=1}^N \omega_j$ with respect to an
inertial frame. In the limit $N
\to \infty$, the quantity $(1/N)\sum_{j=1}^N \omega_j$ coincides with the mean $\langle \omega \rangle$ of the
distribution ${\cal G}(\omega)$, by virtue of the law of large numbers. Note that for asymmetric unimodal frequency
distribution (the case we do not consider in this work), the frequency
with which the centroid turns around in time does not coincide with the mean of the frequency distribution \cite{Basnarkov:2008}.

From Eq.~(\ref{eq:kuramoto-single}), we may easily understand the
tendencies of the two terms on the rhs of the equation in dictating the behavior of the
angles. The first term alone induces every oscillator to oscillate at
its own natural frequency independently of the others, thereby promoting an
unsynchronized state. By contrast, the mean-field term alone promotes
synchrony, as may be seen in the following way. Suppose at some instant
of time $t$, a few of the oscillator angles happen to come close together on the
circle, so that $r(t)$ and $\psi(t)$ have non-zero values. The dynamics
${\rm d}\theta_j/{\rm d}t=-Kr(t)\sin(\theta_j-\psi(t))$, which has a
fixed point at $\theta_j=\psi(t)$, would then tend
to pull the $\theta_j$'s toward the instantaneous average angle
$\psi(t)$. However, the effectiveness with which the $\theta_j$'s are
pulled toward $\psi(t)$ is proportional to the instantaneous amount of
synchrony $r(t)$ present in the system, a feature that leads to a
positive feedback loop being set up between coupling and synchrony: as
more and more oscillators are pulled
toward the instantaneous average angle, the value of $r$, and,
consequently, the effective pull strength $Kr$ grows, which in turn
results in even more oscillators being pulled into the synchronized bunch. The
process continues if further synchrony is promoted by more
oscillators joining the synchronized bunch, or else, the process becomes
self-limiting in time.
The competing tendencies of the natural frequency
and the mean-field term may be best inferred from numerical simulation
results of the dynamics~(\ref{eq:kuramoto-single}) for finite but large
$N$. Simulations for a given unimodal ${\cal G}(\omega)$ reveal
that for values of $K$ less than a critical value $K_c$, the quantity
$r(t)$ while starting from any initial condition decays
at long times to a time-independent value equal to zero, with
fluctuations of $O(N^{-1/2})$. For $K > K_c$, however, $r(t)$ grows
exponentially in time to a time-independent value that is non-zero,
still with fluctuations of order $N^{-1/2}$~\cite{Strogatz:2000}. 

The above-mentioned results make us conclude
that the dynamics~(\ref{eq:kuramoto-single}) leads at long times to a
stationary state in which both $r$ and $\psi$ attain time-independent
values, which we denote by
$r_{\rm st}$ and $\psi_{\rm st}$, respectively. Moreover, for a given
${\cal G}(\omega)$, qualitatively different stationary-state behavior
emerges as $K$ is tuned from small to high values across $K_c$: Small
$K<K_c$ (respectively, large $K>K_c$) promotes an incoherent
(respectively, a synchronized) stationary state. An
unsynchronized/incoherent/homogeneous stationary state implies having the oscillator angles remaining scattered around
randomly on the circle at all times, resulting in the value $r_{\rm
st}=0$. A synchronized stationary state implies having a set of
oscillator angles differing from one
another by time-independent constant values so that the corresponding population moves around the circle in one compact bunch, and one
has $r_{\rm st}>0$. In the case when one has a macroscopic population of
$O(N)$ of synchronized oscillators, we may conclude by invoking the line
of argument mentioned following Eq.~(\ref{eq:kuramoto-model-sumj}) that
the synchronized bunch moves around the unit circle with uniform frequency $\langle \omega \rangle$. In the limit $K \to \infty$, there is only one such synchronized
bunch (thus yielding $r_{\rm st}=1$), while the number of synchronized oscillators steadily decreases
to zero as $K$ decreases towards $K_c$. 

\begin{figure}[ht]
\centering
\includegraphics[width=0.6\textwidth]{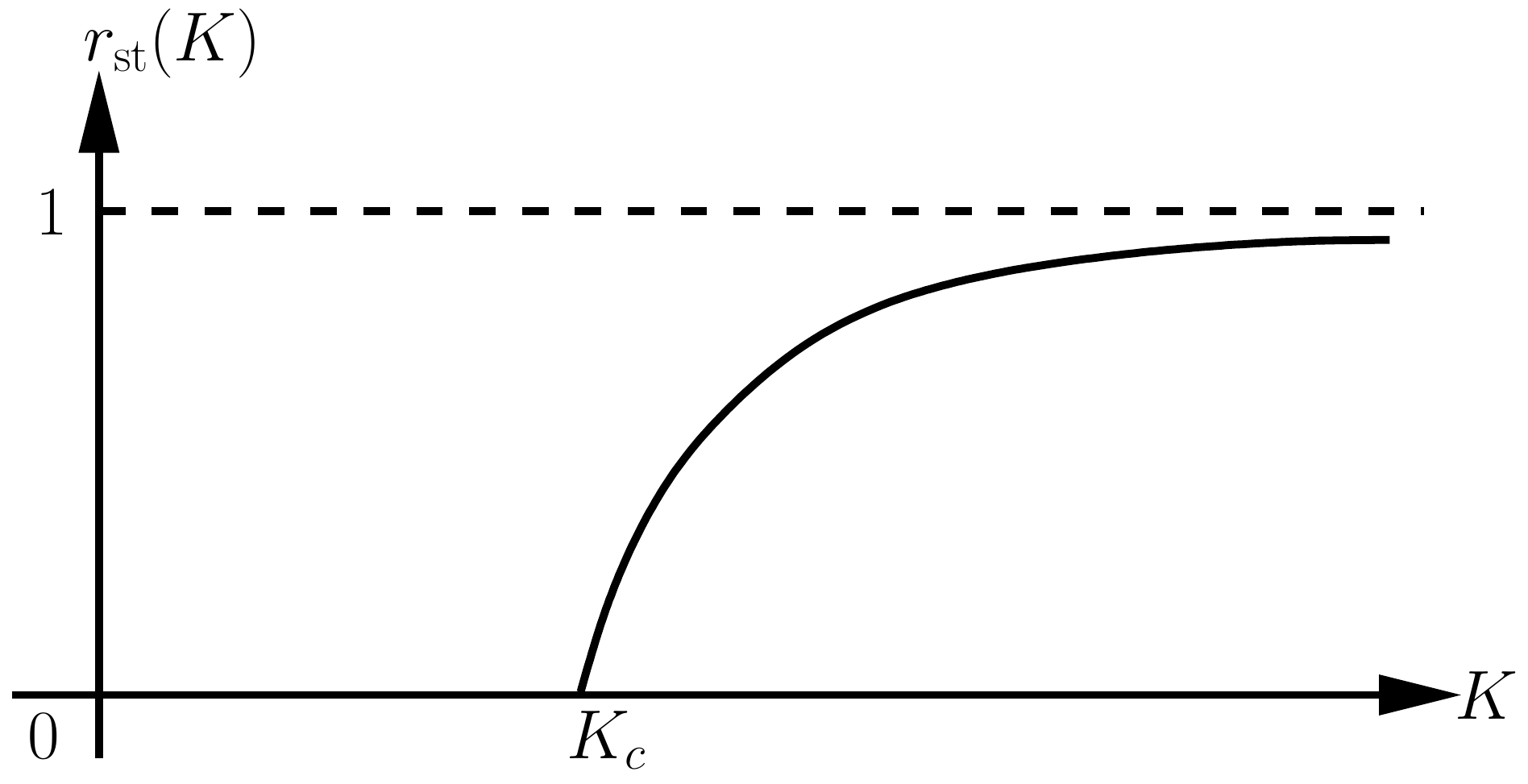}
\caption{The figure shows the schematic dependence of the
stationary-state order parameter $r_{\rm st}$ on
the coupling strength $K$ for the Kuramoto
dynamics~(\ref{eq:kuramoto-single}), with number of oscillators equal to
$N$, and with a
unimodal ${\cal G}(\omega)$ that has a non-compact support. The
figure shows the behavior in the thermodynamic limit
$N\rightarrow\infty$, when it is known that the Kuramoto model admits a second-order phase transition at the critical coupling
strength $K_c$.}
\label{fig:kuramoto-phase-transition}
\end{figure}

Now, in the language of statistical physics, the
observation of qualitatively different macroscopic behaviors on tuning of a control
parameter is referred to as a phase transition, a phenomenon that may be
argued to be possible only in the thermodynamic limit~\cite{Huang:1987}. At a quantitative
level, a phase transition is characterized by different values of the
so-called order parameter, which usually varies between zero in one phase and nonzero in the other. In the context of the Kuramoto model, the
quantity $r_{\rm st}$ plays the role of an order parameter. For $K<K_c$
(respectively, $K > K_c$), one has a homogeneous (respectively, a
synchronized) phase characterized by $r_{\rm st}=0$ (respectively,
$r_{\rm st}>0$). On tuning $K$ across $K_c$, one observes a second-order
or a continuous phase
transition, characterized by a continuous increase of $r_{\rm st}$ from
zero as $K$ is increased beyond $K_c$~\cite{Strogatz:2000,Acebron:2005}. The phase transition in 
the Kuramoto model for a unimodal ${\cal G}(\omega)$ is shown
schematically in
Fig.~\ref{fig:kuramoto-phase-transition}. 

\subsection{Analysis in the thermodynamic limit}
\label{sec:thermodynamic-limit-kuramoto}
In this section, we discuss the analytical properties of the Kuramoto
model (\ref{eq:kuramoto-single})
in the thermodynamic limit $N \to \infty$. Before proceeding, let us note that the effect
of $\langle \omega \rangle$ can be gotten rid of from the
dynamics~(\ref{eq:kuramoto-single}) by viewing the latter in a frame
that is rotating uniformly with frequency $\langle \omega \rangle$ with
respect to an inertial frame; this is tantamount to implementing the
Galilean shift $\theta_j \to \theta_j +\langle \omega \rangle
t~\forall~j$ that leaves the dynamics invariant. In the following, we
will implement such a transformation, and consider from now on the
$\omega_j$'s to be random variables distributed according to the 
distribution $g(\omega)\equiv{\cal G}(\omega+\langle \omega \rangle)$
with zero mean; note that $g(\omega)=g(-\omega)$. 
In this way of looking at things from the rotating frame,
a macroscopic number of oscillators that are synchronized would have
angle points that are immobile on the unit
circle, while oscillators that are out of synchrony would have angle points going
around the unit circle in time. Note that the
dynamics~(\ref{eq:kuramoto-single}) has now only two dynamical parameters: the width $\sigma
\equiv \langle \omega^2 \rangle - \langle \omega \rangle^2$ of
$g(\omega)$, which characterizes how different the individual natural
frequencies are, and the coupling strength $K$, which characterizes how
strongly the oscillators are affecting the motion of each other. 

In the thermodynamic limit, it is natural to characterize the Kuramoto
system in terms of a single-oscillator probability density $\rho(\theta,\omega,t)$ defined
such that $\rho(\theta,\omega,t){\rm d}\theta$ gives the fraction of
oscillators with natural frequency $\omega$ that have their angle lying
between $\theta$ and $\theta+{\rm d}\theta$ at time
$t$~\cite{Strogatz:2000,Acebron:2005}. Note that
invoking the concept of a probability density to describe a collection of
dynamical variables and studying the time evolution of the density due
to the dynamics of the dynamical variables is an approach adopted in
statistical physics to analyze the dynamical behavior of a system. This
approach may be contrasted with the one invoked in dynamical system
theory, where instead one studies the time evolution of individual
dynamical equations (for example, the set of coupled
equations~(\ref{eq:kuramoto-single})) for given initial values of the dynamical
variables.

The density
$\rho$ is non-negative, $2\pi$-periodic in $\theta$, and satisfies the
normalization $\int_{-\pi}^{\pi} {\rm
d}\theta~\rho(\theta,\omega,t)=1~\forall~\omega,t$. Since the total
number of oscillators with a given natural frequency $\omega$ is
conserved by the dynamics~(\ref{eq:kuramoto-single}), the time evolution
of $\rho$ follows a continuity equation that may be derived by
considering a small segment between $\theta$ and $\theta+\Delta \theta$
on the unit circle and oscillators with natural frequency equal to
$\omega$. Then, one may equate the change in a small time
$\Delta t$ the number of oscillator angle points contained in the segment,
given by $\Delta \theta\left[\rho(\theta,\omega,t+\Delta
t)-\rho(\theta,\omega,t)\right]$, with the net number of angle points
that have entered the segment in time $\Delta t$, given by
$\left[J(\theta,\omega,t)-J(\theta+\Delta \theta,\omega,t)\right]\Delta
t$. Here, $J(\theta,\omega,t)=v(\theta,\omega,t)\rho(\theta,\omega,t)$ is
the current at location $\theta$ at time $t$, with $v(\theta,\omega,t)$
being the local velocity at position $\theta$. From
Eq.~(\ref{eq:kuramoto-single}), we have
$v(\theta,\omega,t)=\omega-Kr(t)\sin(\theta-\psi(t))$, where the relation
\begin{equation}
r(t)e^{i\psi(t)}=\int_{-\infty}^\infty {\rm d}\omega\int_{-\pi}^{\pi} {\rm
d}\theta~g(\omega)\rho(\theta,\omega,t)e^{i\theta}\,
\label{eq:rpsi-continuum}
\end{equation}
is obtained as the
$N\to \infty$ generalization of Eq.~(\ref{eq:rpsi-definition}).
Realizing that the equality $\Delta \theta\left[\rho(\theta,\omega,t+\Delta
t)-\rho(\theta,\omega,t)\right]=\left[J(\theta,\omega,t)-J(\theta+\Delta \theta,\omega,t)\right]\Delta
t$ holds for arbitrary
$\Delta \theta$, we get in the limit $\Delta t \to 0$ the continuity equation $\partial
\rho(\theta,\omega,t)/\partial t+\partial
\left(v(\theta,\omega,t)\rho(\theta,\omega,t)\right)/\partial \theta=0$,
i.e.,
\begin{equation}
\frac{\partial \rho}{\partial
t}+\frac{\partial}{\partial
\theta}\left[\left(\omega+K\int_{-\infty}^\infty \int_{-\pi}^{\pi} {\rm
d}\theta'~\sin(\theta'-\theta)g(\omega)\rho(\theta',\omega,t)\right)\rho\right]=0\,. 
\label{eq:continuity-equation}
\end{equation}

A stationary state of the dynamics~(\ref{eq:kuramoto-single}) would mean
to have a density such that $\partial \rho(\theta,\omega,t)/\partial t=0$, that is, a time-independent density $\rho_{\rm
st}(\theta,\omega)$ that satisfies
\begin{equation}
\frac{\partial}{\partial
\theta}\left[\left(\omega+K\int_{-\infty}^\infty \int_{-\pi}^{\pi} {\rm
d}\theta'~\sin(\theta'-\theta)g(\omega)\rho_{\rm st}(\theta',\omega)\right)\rho_{\rm st}(\theta,\omega)\right]=0\,. 
\label{eq:continuity-equation-stationary}
\end{equation}
Note that a state is to be considered stationary only in the
statistical sense: in such a state, although individual oscillators continue to change
their angles in accordance with the dynamics
\begin{equation}
\frac{{\rm d}\theta_j}{{\rm d}t}=\omega_j-Kr_{\rm
st}\sin(\theta_j-\psi_{\rm st})\,,
\label{eq:kuramoto-single-stationary}
\end{equation}
the number of oscillators with a given value of the angle is constant in
time. In
Eq.~(\ref{eq:kuramoto-single-stationary}), we may set $\psi_{\rm st}$ to
zero by choosing suitably the origin of the angle axis, see
Fig.~\ref{fig:rpsi}. Such a choice
would correspond to having the stationary values $r_{y,{\rm st}}=0$ and
$r_{x,{\rm st}}=r_{\rm st}$, see Eq.~(\ref{eq:rxry}). Consequently,
one has
\begin{equation}
r_{\rm st}=\int_{-\infty}^\infty {\rm d}\omega\int_{-\pi}^{\pi} {\rm
d}\theta~g(\omega)\rho_{\rm st}(\theta,\omega)\cos \theta\,.
\label{eq:kuramoto-rst}
\end{equation}

In his early works, Kuramoto adduced a remarkable analysis to predict the critical value $K_c$ such that $r_{\rm st}=0$
for $K \le K_c$ and $r_{\rm st}>0$ for $K > K_c$~\cite{Kuramoto:1975}. The analysis cleverly
bypasses the
formidable task of solving explicitly Eq.~(\ref{eq:continuity-equation-stationary}). His prediction for $K_c$, borne out by later
investigations, was
\begin{equation} 
K_c=\frac{2}{\pi g(0)}\,.
\label{eq:kuramoto-Kc}
\end{equation}
We now recall the analysis due to
Kuramoto~\cite{Kuramoto:1975,Strogatz:2000}, which relies on 
adopting the following strategy well-known from statistical mechanical
treatment of mean-field models~\cite{Huang:1987}. At a fixed $K$, we first assume a given value
of $r_{\rm st}$, then (i) obtain the stationary density $\rho_{\rm st}(\theta,\omega)$ implied by the stationary-state dynamics
(\ref{eq:kuramoto-single-stationary}), and finally, (ii) require that the
obtained density when substituted in Eq.~(\ref{eq:kuramoto-rst}) reproduces the given value
of $r_{\rm st}$, thereby yielding a self-consistent equation. 

For a given value of $r_{\rm st}$, it
follows from Eq.~(\ref{eq:kuramoto-single-stationary}) (with $\psi_{\rm
st}=0$) that the dynamics
of oscillators with $|\omega_j| \le Kr_{\rm st}$ approaches in time a stable fixed point
defined by $\omega_j=Kr_{\rm st}\sin \theta_j$, so that the
$j$-th oscillator in this group has after evolving for
a short time a time-independent angle given by
$\theta_j=\sin^{-1}[\omega_j/(Kr_{\rm st})]$; $|\theta_j| \le
\pi/2$. This group of oscillators is thus ``locked'' or
synchronized, and has the density
\begin{equation}
\rho_{\rm st}(\theta,\omega)=K r_{\rm st} \cos \theta
~\delta\Big(\omega- K r_{\rm st} \sin \theta\Big)
\Theta (\cos \theta ) ;
~~|\omega| \le Kr_{\rm st}\,,
\label{eq:kuramoto-distribution-locked}
\end{equation}
where $\Theta(\cdot)$ is the Heaviside step function, and the prefactor
is derived by the normalization condition
$\int_{-\pi}^{\pi} {\rm d}\theta~ \rho_{\rm st}(\theta,\omega)=1$.
Equation~(\ref{eq:kuramoto-single-stationary}) implies that oscillators
with $|\omega_j| \ge Kr_{\rm st}$ would however have
ever drifting time-dependent angles. On the unit circle, the
corresponding angle points would be buzzing around the circle, spending naturally longer duration at locations that allow for a
smaller local velocity $v(\theta,\omega,t)$ and zipping through locations that have a larger local velocity.
Consequently, the density of this group of ``drifting" oscillators would for most times be
peaked around locations with small local velocities, thus
leading to a stationary density for this group that is inversely proportional to the local velocity:
\begin{equation}
\rho_{\rm st}(\theta,\omega)=\frac{C}{|\omega- K r_{\rm st} \sin
\theta|};
~~|\omega| > Kr_{\rm st}\,.
\label{eq:kuramoto-distribution-drifting}
\end{equation}
Using the normalization condition
$\int_{-\pi}^\pi {\rm d}\theta~ \rho_{\rm st}(\theta,\omega)=1$, we get $C=(1/(2\pi))\sqrt{\omega^2-(Kr_{\rm st})^2}$.

We now require that the given value of $r_{\rm st}$ coincides with the
one implied by Eq.~(\ref{eq:kuramoto-rst}) and the densities in Eqs.~(\ref{eq:kuramoto-distribution-locked}) and
(\ref{eq:kuramoto-distribution-drifting}). Plugging the
latter forms in Eq.~(\ref{eq:kuramoto-rst}), we get
\begin{eqnarray}
&&r_{\rm st}=\int_{-\pi}^\pi {\rm d}\theta \int_{|\omega| > Kr_{\rm st}}
{\rm d}\omega~g(\omega)\frac{C\cos \theta}{|\omega- K r_{\rm st} \sin
\theta|}\nonumber \\
&&+\int_{-\frac{\pi}{2}}^{\frac{\pi}{2}} {\rm d}\theta
\int_{|\omega| \le Kr_{\rm st}} {\rm d}\omega~g(\omega) \cos \theta K
r_{\rm st} \cos \theta \,\delta\Big(\omega- K r_{\rm st} \sin
\theta\Big)\,.
\end{eqnarray}
The first integral on the rhs vanishes due to the symmetry
$g(\omega)=g(-\omega)$ and the property that $\rho_{\rm
st}(\theta+\pi,-\omega)=\rho_{\rm st}(\theta,\omega)$ for
the group of drifting oscillators, as given by Eq.~(\ref{eq:kuramoto-distribution-drifting}). The second integral after integration over $\omega$ yields
the desired self-consistent equation
\begin{equation}
r_{\rm st}=Kr_{\rm st}\int_{-\pi/2}^{\pi/2}{\rm d} \theta~\cos^2 \theta ~g(Kr_{\rm st}\sin
\theta)\,.
\end{equation}
This equation has the 
trivial solution $r_{\rm st} = 0$ valid for any $K$, which corresponds
to the incoherent state with density $\rho_{\rm st}^{\rm inc}(\theta,\omega) =
1/(2\pi) ~\forall~ \theta,\omega$. One also has a solution with $r_{\rm st} \ne 0$ that satisfies
\begin{equation}
1=K\int_{-\pi/2}^{\pi/2}{\rm d} \theta~ \cos^2 \theta ~g(Kr_{\rm st}\sin
\theta)\,,
\label{eq:kuramoto-bifurcation}
\end{equation}
which bifurcates continuously from the incoherent solution at
the value $K = K_c$ obtained from
the above equation on taking the limit $r_{\rm st} \to 0^+$. Since for a
unimodal $g(\omega)$, one has a negative second derivative at $\omega = 0$, i.e., $g''(0)<0$, one finds by expanding
the integrand in Eq.~(\ref{eq:kuramoto-bifurcation}) as a powers series in $r_{\rm st}$ that the
bifurcation in this case is supercritical. It may be shown that
consistently with Fig.~\ref{fig:kuramoto-phase-transition}, a solution
$r_{\rm st}$ of Eq.~(\ref{eq:kuramoto-bifurcation})
exists for $K\ge K_c$, which equals $0$ for $K=K_c$, and which
increases with $K$ and approaches unity as $K\to
\infty$~\cite{Gupta:2014-2}.

The linear stability of the incoherent state $\rho_{\rm st}^{\rm
inc}$ may be studied by expanding $\rho(\theta,\omega,t)$ as
$\rho(\theta,\omega,t)=\rho_{\rm st}^{\rm inc}(\theta,\omega)+\epsilon
e^{\lambda t}\delta \rho(\theta,\omega);~~|\epsilon| \ll
1$~\cite{Strogatz:2000}. Here, the
parameter $\lambda$ determines the stability properties of the
incoherent state: when $\lambda$ has a positive (respectively, a
negative) real part, the state is linearly stable (respectively,
unstable), while a purely imaginary $\lambda$ implies that the state is
linearly neutrally stable. Further, noting that $\rho(\theta,\omega,t)$,
and hence, $\delta \rho(\theta,\omega)$ is $2\pi$-periodic in
$\theta$, a Fourier expansion yields $\delta
\rho(\theta,\omega)=\left(\widetilde{\delta
\rho}(\omega)e^{i\theta}+{\rm c. c.}\right)+\delta
\rho^{\bot}(\theta,\omega)$, where ${\rm c.c.}$ stands for complex
conjugate, while $\delta \rho^{\bot}(\theta,\omega)$ contains second
and higher harmonics of $\theta$. Substituting in
Eq.~(\ref{eq:continuity-equation}), one obtains an equation linear in
$\widetilde{\delta \rho}(\omega)$, as $\widetilde{\delta
\rho}(\omega)=\left[K/(2(\lambda+i\omega))\right]\int_{-\infty}^\infty {\rm d}\omega'~\widetilde{\delta
\rho}(\omega')g(\omega')$. Multiplying both sides by $g(\omega)$, and
then integrating over $\omega$, one obtains the characteristic equation
determining $\lambda$, as $1=(K/2)\int_{-\infty}^\infty {\rm
d}\omega~g(\omega)/(\lambda+i\omega)$. For our choice of $g(\omega)$
that is even in $\omega$ and nowhere increasing on $\omega \in
[0,\infty)$, it may be shown that the characteristic equation has at
most one solution for $\lambda$, which when it exists is necessarily
real~\cite{Strogatz:2000}. The characteristic equation consequently reads $1=(K/2)\int_{-\infty}^\infty {\rm
d}\omega~\lambda g(\omega)/(\lambda^2+\omega^2)$, which implies that
$\lambda$ can never be negative, and hence, that the incoherent
stationary state can never be linearly stable but is either neutrally
stable or unstable ! The boundary between the neutrally stable and the
unstable behavior is obtained by letting $\lambda\to 0^+$ in the
characteristic equation, thereby yielding the critical value $K_c$ of
Eq.~(\ref{eq:kuramoto-Kc}), such that the state is neutrally stable
(respectively, stable) for $K<K_c$ (respectively, for $K > K_c$). In the
light of the fact that $r(t)$ is obtained as an integral over
$\rho(\theta,\omega,t)$, see Eq.~(\ref{eq:rpsi-continuum}), the
latter fact seems apparently inconsistent with the numerical observation
mentioned previously that for $K<K_c$, the quantity $r(t)$
while starting from any initial condition decays at long times to a time-independent value equal to
zero. Indeed, neutral stability of the incoherent state implies sustained
oscillations of $r(t)$, and whose decay in time, as observed in
simulations, is possible only if a damping mechanism is present in the
dynamics of $r(t)$. It has been rather rigorously demonstrated that
indeed such a mechanism is present as regards the time evolution of
$r(t)$ that draws analogy, as far as its mathematical structure is
concerned, with the phenomenon of Landau damping present
in plasma systems. We refer the reader to Ref.~\cite{Strogatz:2000} for a highly readable account of the phenomenon and its observation
in the Kuramoto model.

\subsection{Noisy Kuramoto model}
\label{sec:noisy-kuramoto}

A rather interesting generalization of the Kuramoto model was studied by
Sakaguchi, who considered the situation in which the Kuramoto oscillators
do not have natural frequencies that are constant in time but which undergo
rapid stochastic fluctuations in time~\cite{Sakaguchi:1988}. Thus, in this model, the natural frequency of
the $j$-th oscillator is a random variable that varies in time (thus
representing annealed disorder) about the
average given by $\omega_j$. Note that in the case of the noisy
Kuramoto model, there are two sources of randomness and two kinds of
averaging involved. The natural frequency of the $j$-th oscillator is an
annealed-disordered random variable that fluctuates in time, with the
time-average denoted by $\omega_j$. The set $\{\omega_j\}_{1\le j \le
N}$, referring to the time-averaged natural frequency of all the oscillators, themselves
represent a set of quenched-disordered random variables sampled from the
distribution $g(\omega)$. As discussed previously, $g(\omega)$ is unimodal
and symmetric about zero, and moreover, decreases monotonically and continuously to zero
with increasing $|\omega|$. The governing equations of motion of the
noisy Kuramoto model are~\cite{Sakaguchi:1988}
\begin{equation}
\frac{{\rm d}\theta_j}{{\rm
d}t}=\omega_j-Kr(t)\sin(\theta_j-\psi(t))+\eta_j(t)\,,
\label{eq:kuramoto-single-noise}
\end{equation}
where $\eta_j(t)$ is a Gaussian, white noise satisfying
\begin{equation}
\langle \eta_j(t) \rangle=0\,, ~~\langle \eta_j(t)\eta_k(t')
\rangle=2D\delta_{jk}\delta(t-t')\,,
\end{equation}
where $D \ge 0$ is a parameter that characterizes noise strength. Here and in the following, we will use angular brackets to denote
averaging over noise realizations. 

\begin{figure}[ht]
 \centering
 \includegraphics[width=0.5\textwidth]{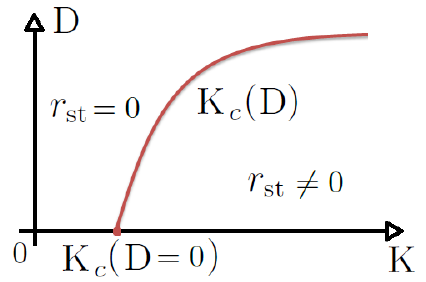}
 \caption{For the noisy Kuramoto model~(\ref{eq:kuramoto-single-noise}),
 the figure shows the phase boundary given by $K_c(D)$ between the
 homogeneous ($r_{\rm st}=0$) and the synchronized ($r_{\rm st}>0$)
 phase, with $K_c(D)$ given by
Eq.~(\ref{eq:KcD-kuramoto-noise}).}
 \label{fig:kuramoto-noise-phase-diagram}
\end{figure}

Note that Eq.~(\ref{eq:kuramoto-single-noise}), which is a stochastic differential equation, has the form of a Langevin equation.
The reader may recall that a Langevin equation describes the time
evolution of a subset of degrees of freedom that are changing only
slowly in comparison to the remaining degrees of freedom of a system~\cite{Gardiner:1983}.
In our case of coupled oscillators, we take the natural frequencies of
the oscillators to
be fluctuating about their average values on a much faster timescale than the
one over which the angle $\theta_j$'s are evolving, and it is the former fast variation that leads to the
stochastic noise $\eta_j(t)$ in the equations of motion.
Equation~(\ref{eq:kuramoto-single-noise}) being a representative
Langevin dynamics may be studied by employing the corresponding tool of
analysis usual in statistical physical studies, namely, the
Fokker-Planck equation~\cite{Gardiner:1983,Risken:1996} for the time
evolution of the single-oscillator probability density
$\rho(\theta,\omega,t)$ defined above. This equation may be derived
straightforwardly for the dynamics~(\ref{eq:kuramoto-single-noise}), and
has the form
\begin{equation}
\frac{\partial \rho}{\partial t}=D\frac{\partial^2\rho}{\partial
\theta^2}-\frac{\partial}{\partial
\theta}\left[\left(\omega+K\int_{-\infty}^\infty \int_{-\pi}^{\pi} {\rm
d}\theta'~\sin(\theta'-\theta)g(\omega)\rho(\theta',\omega,t)\right)\rho\right]=0\,.
\label{eq:Fokker-Planck-kuramoto-noise}
\end{equation}
For $D=0$, the above equation reduces to the continuity equation
of the Kuramoto model, Eq.~(\ref{eq:continuity-equation}), as it should.
Sakaguchi extended the self-consistent analysis of the Kuramoto
model presented above to address the issue of which critical value of $K$ allows in the stationary state
for a branch of synchronized states to bifurcate from an incoherent
state. The critical value is obtained as~\cite{Sakaguchi:1988,Gupta:2014-2}
\begin{equation}
K_c(D)=2\left[\int_{-\infty}^{+\infty} {\rm
d}\omega~\frac{g(D\omega)}{\omega^2+1}\right]^{-1}\,,
\label{eq:KcD-kuramoto-noise}
\end{equation}
which as $D \to 0^+$ may be checked to correctly reduce to the expected
answer, namely, $K_c(0^+)$ equals $K_c$ given by Eq.~(\ref{eq:kuramoto-Kc}).
It may be shown that the incoherent stationary state $\rho_{\rm
st}^{\rm inc}(\theta,\omega)=1/(2\pi)$ is linearly stable under the
dynamics~(\ref{eq:Fokker-Planck-kuramoto-noise}) for $K < K_c(D)$ and is
linearly unstable for $K > K_c(D)$. Consequently, for $K<K_c(D)$
(respectively, $K > K_c(D($), one has a homogeneous (respectively, a
synchronized) phase characterized by $r_{\rm st}=0$ (respectively,
$r_{\rm st}>0$). On tuning $K$, one observes a continuous transition
between the two phases at $K=K_c(D)$. Figure~\ref{fig:kuramoto-noise-phase-diagram} shows the phase boundary given by $K_c(D)$ between the
homogeneous and the synchronized phase.

\section{Generalized Kuramoto model with inertia and noise}
\label{sect:kuramoto-first-order}
In this section, we study a very interesting generalization of the
Kuramoto dynamics (\ref{eq:kuramoto-single}) that includes inertial terms
parametrized by a moment
of inertia and stochastic noise, as discussed in
Refs.~\cite{Tanaka:1997,Acebron:1998,Hong:1999,Acebron:2000,Gupta:2014,Gupta:2014-2}. Inclusion of inertia
elevates the first-order Kuramoto dynamics to one that is second order
in time, while noise accounts for temporal fluctuations
of the natural frequencies. The generalization offers the possibility to
explore the issue of emergence of spontaneous synchronization in a wider
space of parameters, and, as we will discuss below, leads even with a
unimodal natural frequency distribution to a rather rich phase diagram
relative to the Kuramoto model that includes both equilibrium and nonequilibrium phase transitions.
Besides, the generalized model represents a bridge between two
apparently disconnected research areas, namely, the area of spontaneous
synchronization pursued by dynamical physicists and that of statistical
physical studies, in both in and out of equilibrium regimes, of so-called long-range
interacting systems pursued within the community of statistical
physicists. It turns out that two different limits of the
generalized model have been studied extensively over the years, albeit
with not much overlap and inter-community dialogue, by the communities
of dynamical and statistical physicists. 

In the generalized dynamics, a dynamical variable in addition to the
angle $\theta_j$, namely, angular
velocity $v_j$, is assigned to each oscillator, so that the equations of
motion are~\cite{Acebron:1998,Hong:1999,Acebron:2000}:
\begin{equation}
\frac{{\rm d}\theta_j}{{\rm d} t}=v_j\,,~~m\frac{{\rm d} v_j}{{\rm d}
t}=-\gamma v_j+\gamma\omega_j-\widetilde{K}
r\sin(\theta_j-\psi)+\widetilde{\eta}_j(t)\,. 
\label{eq:first-order-eom} 
\end{equation}
Here, $m$ is the common moment of inertia of the oscillators, $\gamma>0$
is a parameter that plays the role of a damping constant, $\widetilde{K}$ is the strength of coupling
between the oscillators, 
while $\widetilde{\eta}_j(t)$ is a Gaussian, white noise satisfying
\begin{equation}
\langle \widetilde{\eta}_j(t) \rangle=0\,, ~~\langle
\widetilde{\eta}_j(t)\widetilde{\eta}_k(t')
\rangle=2\widetilde{D}\delta_{jk}\delta(t-t')\,.
\label{eq:etatilde-prop}
\end{equation}
Here, $\widetilde{D}\ge0$ is a parameter that sets the strength of the
noise.

That $\gamma$ plays the role of a damping constant in the
dynamics~(\ref{eq:first-order-eom}) may be appreciated by considering
the noise-average of the second equation in~(\ref{eq:first-order-eom})
that yields the dynamics $m{\rm d} \langle v_j \rangle/{\rm d}
t=-\gamma \langle v_j\rangle+\gamma\omega_j-\widetilde{K}
\langle r\sin(\theta_j-\psi)\rangle$, which shows that in the absence of natural frequencies and the
interaction between the oscillators, any average initial velocity decays
to zero (natural frequencies and interaction would of course not let this happen!).  

It is worth noting that the dynamics~(\ref{eq:first-order-eom}) without
the noise term, studied in~\cite{Tanaka:1997}, arises in a completely different context, namely, in electrical
power distribution networks comprising synchronous
generators (representing power plants) and motors (representing
customers)~\cite{Filatrella:2008,Rohden:2012}; the dynamics arises in the approximation in which every node of the network is
connected to every other.

In the limit of overdamped motion ($m \to 0$ at a fixed $\gamma \ne 0$),
the dynamics~(\ref{eq:first-order-eom}) reduces to
\begin{equation}
\gamma \frac{{\rm d}\theta_j}{{\rm d} t}=\gamma \omega_j-\widetilde{K}
r\sin(\theta_j-\psi)+\widetilde{\eta}_j(t)\,.
\label{eq:eom-overdamped1}
\end{equation}
Then, defining $K \equiv \widetilde{K}/\gamma$ and
$\eta_j(t) \equiv \widetilde{\eta}_j(t)/\gamma$ so that
$D=\widetilde{D}/\gamma^2$, the dynamics (\ref{eq:eom-overdamped1}) for $D=0$
becomes that of the Kuramoto model, Eq.~(\ref{eq:kuramoto-single}), and for $D
\ne 0$ that of its noisy version given by the
dynamics~(\ref{eq:kuramoto-single-noise}).

\subsection{The model as a long-range interacting
system}
\label{sec:longrangemodel}
It may be shown that in a different context than that of coupled oscillators,
the dynamics~(\ref{eq:first-order-eom}) describes a long-range
interacting system of particles moving on a unit circle, with each
particle acted upon by a quenched external torque $\widetilde{\omega}_j \equiv \gamma
\omega_j$. Recent exploration of long-range interacting systems, and in
particular, of their static and dynamic properties, has focussed on an
analytically tractable and representative model called the Hamiltonian
mean-field (HMF) model~\cite{Ruffo:1995,Inagaki:1993}.

Long-range interacting (LRI) systems are those in which the
inter-particle interaction potential decays slower than $1/r^d$, with
$d$ being the dimension of the embedding space~\cite{Campa:2009,Gupta:2010,Campa:2014,Levin:2014,Gupta:2017}. Unlike
short-range ones, LRI systems are intrinsically nonadditive, namely,
they cannot be trivially divided into independent macroscopic subparts. LRI systems are quite ubiquitous in
Nature, typical examples being self-gravitating systems, charged
plasmas, two-dimensional quasi-geostrophic flows, wave-particle
interaction in plasma, etc. The feature
of nonadditivity of LRI systems leads to many fascinating phenomena not exhibited by
short-range systems, such as
inequivalence of statistical ensembles, breaking of ergodicity,
occurrence of long-lived non-Boltzmann quasistationary states during
relaxation to equilibrium, etc~\cite{Campa:2014,Gupta:2017}. 

The HMF model comprises $N$ particles of mass
$m$ moving on a unit circle and interacting through a long-range interparticle
potential that is of the mean-field type: every particle is coupled to every other with equal strength.
The Hamiltonian of the HMF model is~\cite{Ruffo:1995}
\begin{equation}
H=\sum_{j=1}^{N}\frac{p_j^2}{2m}+\frac{\widetilde{K}}{2N}\sum_{j,k=1}^{N}\left[1-\cos(\theta_j-\theta_k)\right]\,,
\label{eq:HMF-H}
\end{equation}
where $\theta_j \in [-\pi,\pi]$ gives the position of the $j$-th particle
on the circle, while $p_j=mv_j$ is its conjugated angular
momentum, with $v_j$ being the angular velocity. 
The time evolution of the system within a microcanonical ensemble
follows the deterministic Hamilton equations of motion: 
\begin{equation}
\frac{{\rm d} \theta_j}{{\rm d} t}=v_j\,,~~m\frac{{\rm d} v_j}{{\rm d}
t}=-\widetilde{K}
r\sin(\theta_j-\psi)\,.
\label{eq:hameq} 
\end{equation}
The dynamics conserves the total energy and momentum, and leads at long times to
an equilibrium stationary state in which, depending on the energy
density $\epsilon \equiv H/N$, the system could be in one of two possible
phases: for $\epsilon$ smaller than a critical value $\epsilon_c=3\widetilde{K}/4$, the
system is in a clustered phase in which the particles are close together
on the circle, while for $\epsilon > \epsilon_c$, the particles
are uniformly distributed on the circle, thus characterizing a homogeneous
phase~\cite{Campa:2009}. A continuous phase transition between the two
phases is characterized by a positive value of $r_{\rm st}$ in the clustered
phase and a zero value in the homogeneous phase.

One may generalize the microcanonical dynamics (\ref{eq:hameq}) to
account for interaction with an external heat bath at temperature
$T$. The resulting model, called the Brownian mean-field (BMF) model, has thus a canonical ensemble dynamics
given by~\cite{Chavanis:2014}.
\begin{equation}
\frac{{\rm d} \theta_j}{{\rm d} t}=v_j\,,~~m\frac{{\rm d} v_j}{{\rm d}
t}=-\gamma v_j-\widetilde{K}
r\sin(\theta_j-\psi)+\widetilde{\eta}_j(t)\,, 
\label{eq:bmfeq}
\end{equation}
where $\widetilde{\eta}_j(t)$ is as in Eq.~(\ref{eq:etatilde-prop}). One may
then invoke the fluctuation-dissipation relation to express the strength $\widetilde{D}$
of the noise in terms of the temperature $T$ and the damping constant
$\gamma$ as $\widetilde{D}=\gamma k_BT$~\cite{Huang:2009}.
We will set the Boltzmann constant $k_B$ to unity in the rest of the paper. 
The canonical dynamics~(\ref{eq:bmfeq}) also leads to a long-time
equilibrium stationary state in which a generic configuration $C\equiv
\{\theta_j,v_j\}_{1\le j \le N}$ with energy
$E(C)$ occurs with the usual Gibbs-Boltzmann weight: $P_{\rm eq}(C)
\propto \exp[-E(C)/T]$. 
The phase transition in the HMF model observed within the
microcanonical ensemble now occurs within the canonical ensemble as one
tunes the temperature across the critical value $T_c=\widetilde{K}/2$.
The derivation of this result is discussed below, namely, in
Section~\ref{sec:BMF-stationary-state}.

Let us now consider a set of quenched external
torques $\{\widetilde{\omega}_j\equiv \gamma \omega_j\}$ acting on each of the particles,
thereby pumping energy into the system. In this case, the second
equation in the canonical dynamics~(\ref{eq:bmfeq}) has an additional
term $\widetilde{\omega}_j$ on
the rhs. The resulting dynamics becomes exactly the same as the
dynamics~(\ref{eq:first-order-eom}) of the generalized Kuramoto model. 

\subsection{Dynamics in a reduced parameter space}
\label{secreducedspace}
It proves convenient to reduce the number of parameters in the
dynamics~(\ref{eq:first-order-eom}). To this end, we note that the
effect of $\sigma$ may be made explicit by replacing $\omega_j$
in the second equation by $\sigma \omega_j$. Therefore, we will consider from now
on the dynamics~(\ref{eq:first-order-eom}) with
the substitution $\omega_j \rightarrow \sigma \omega_j$. In the resulting
model, $g(\omega)$ therefore has zero mean and unit width. Moreover, we will
consider in the dynamics (\ref{eq:first-order-eom}) the parameter $\widetilde{D}$ to be
$\widetilde{D}=\gamma T$, a relation we discussed above.

For $m \ne 0$, using dimensionless quantities~\cite{Gupta:2014,Gupta:2014-2} 
\begin{equation}
\overline{t}\equiv
t\sqrt{\widetilde{K}/m},~\overline{v}_j\equiv
v_j\sqrt{m/\widetilde{K}},~1/\sqrt{\overline{m}}\equiv
\gamma/\sqrt{\widetilde{K}m},~\overline{\sigma} \equiv
\gamma \sigma/\widetilde{K},~\overline{T} \equiv
T/\widetilde{K},~\overline{\eta}_j(\overline{t})\equiv
\widetilde{\eta}_j(t)/\widetilde{K}\,,
\label{eq:dimensionless-definition}
\end{equation}
the equations of motion~(\ref{eq:first-order-eom}) become
\begin{equation}
\frac{{\rm d}\theta_j}{{\rm
d}\overline{t}}=\overline{v}_j\,,~~\frac{{\rm d}\overline{v}_j}{{\rm
d}\overline{t}}=-\frac{1}{\sqrt{\overline{m}}}\overline{v}_j-r\sin(\theta_j-\psi)+\overline{\sigma}\omega_j+\overline{\eta}_j(\overline{t})\,,
\label{eq:eom-scaled}
\end{equation}
where $\langle \overline{\eta}_j(\overline{t})\overline{\eta}_k(\overline{t}')
\rangle=2(\overline{T}/\sqrt{\overline{m}})\delta_{jk}\delta(\overline{t}-\overline{t}')$.
For $m=0$, using dimensionless time $\overline{t}\equiv
t(\widetilde{K}/\gamma)$, with $\overline{\sigma}$ and $\overline{T}$ as defined above, the dynamics becomes the overdamped motion
\begin{equation}
\frac{{\rm d}\theta_j}{{\rm
d}\overline{t}}=\overline{\sigma}\omega_j-r\sin(\theta_j-\psi)+\overline{\eta}_j(\overline{t})\,,
\label{eq:eom-overdamped}
\end{equation}
where we have $\langle \overline{\eta}_j(\overline{t})\overline{\eta}_k(\overline{t}')
\rangle=2\overline{T}\delta_{jk}\delta(\overline{t}-\overline{t}')$.
We thus have in place of the dynamics
(\ref{eq:first-order-eom}) involving five parameters,
$m,\gamma,\widetilde{K},\sigma,T$ the reduced dynamics
(\ref{eq:eom-scaled}) (or (\ref{eq:eom-overdamped}) in
the overdamped limit) that involves three
dimensionless parameters, $\overline{m},\overline{T},\overline{\sigma}$.
We will from now on consider the dynamics in this reduced parameter space, dropping 
overbars for simplicity of notation.
With $\sigma=0$ (i.e. $g(\omega)=\delta(\omega)$ 
the dynamics~(\ref{eq:eom-scaled}) is that of the BMF model with an equilibrium stationary
state. For other $g(\omega)$, it may be shown that the
dynamics~(\ref{eq:eom-scaled}) violates detailed balance, leading to a
NESS~\cite{Gupta:2014}. 

\subsection{Nonequilibrium first-order synchronization phase
transition}
\label{sec:phasetransition}
In this section, we report results on a very interesting nonequilibrium phase
transition that occurs in the stationary state of the dynamics
(\ref{eq:eom-scaled}). In the three-dimensional space of parameters
$(m,T,\sigma)$, let us
first locate the phase transitions in the Kuramoto model,
Eq.~(\ref{eq:kuramoto-single}), and in its noisy extension,
Eq.~(\ref{eq:kuramoto-single-noise}),
respectively. 
\begin{itemize}
\item{The phase transition of the Kuramoto dynamics ($m=T=0$, $\sigma
\ne 0$) corresponds to a continuous transition from a
low-$\sigma$ synchronized to a high-$\sigma$ incoherent
 phase across the critical point 
 \begin{equation}
 \sigma_c(m=0,T=0)=\frac{\pi g(0)}{2}\,,
 \end{equation}
 which is obtained using Eq.~(\ref{eq:kuramoto-Kc}), see Ref.~\cite{Gupta:2014-2}.}
 \item{Extending the Kuramoto dynamics to $T \ne 0$ (the noisy Kuramoto
 model), the aforementioned critical point becomes a second-order critical line on the $(T,\sigma)$-plane, given by solving
 \begin{equation}
2=\int_{-\infty}^{\infty}d\omega ~\frac{g(\omega)
T}{T^2+\omega^2\sigma^2_c(m=0,T))}\,.
\label{eq:sakaguchi-again}
\end{equation}
The above equation is obtained by using
Eq.~(\ref{eq:KcD-kuramoto-noise}), as may be looked up in Ref.~\cite{Gupta:2014-2}.}
\item{The transition in the BMF dynamics ($m,T \ne 0, \sigma=0$)
corresponds now to a continuous transition occurring at the critical
temperature $T_c=1/2$. This result in proved in
Section~\ref{sec:BMF-stationary-state}.}
\end{itemize}

\begin{figure}[!ht]
\centering
\includegraphics[width=0.5\textwidth]{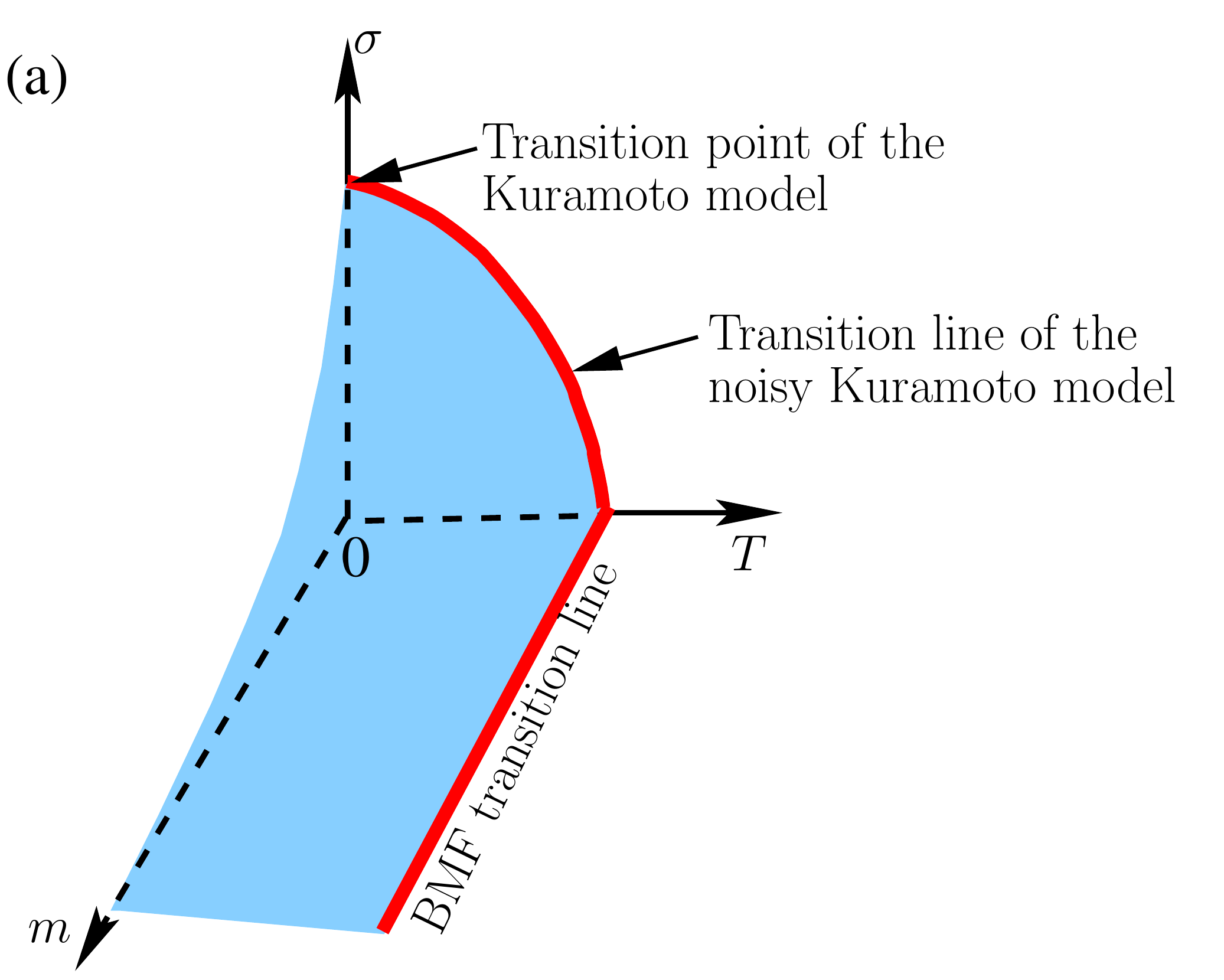}
\includegraphics[width=0.5\textwidth]{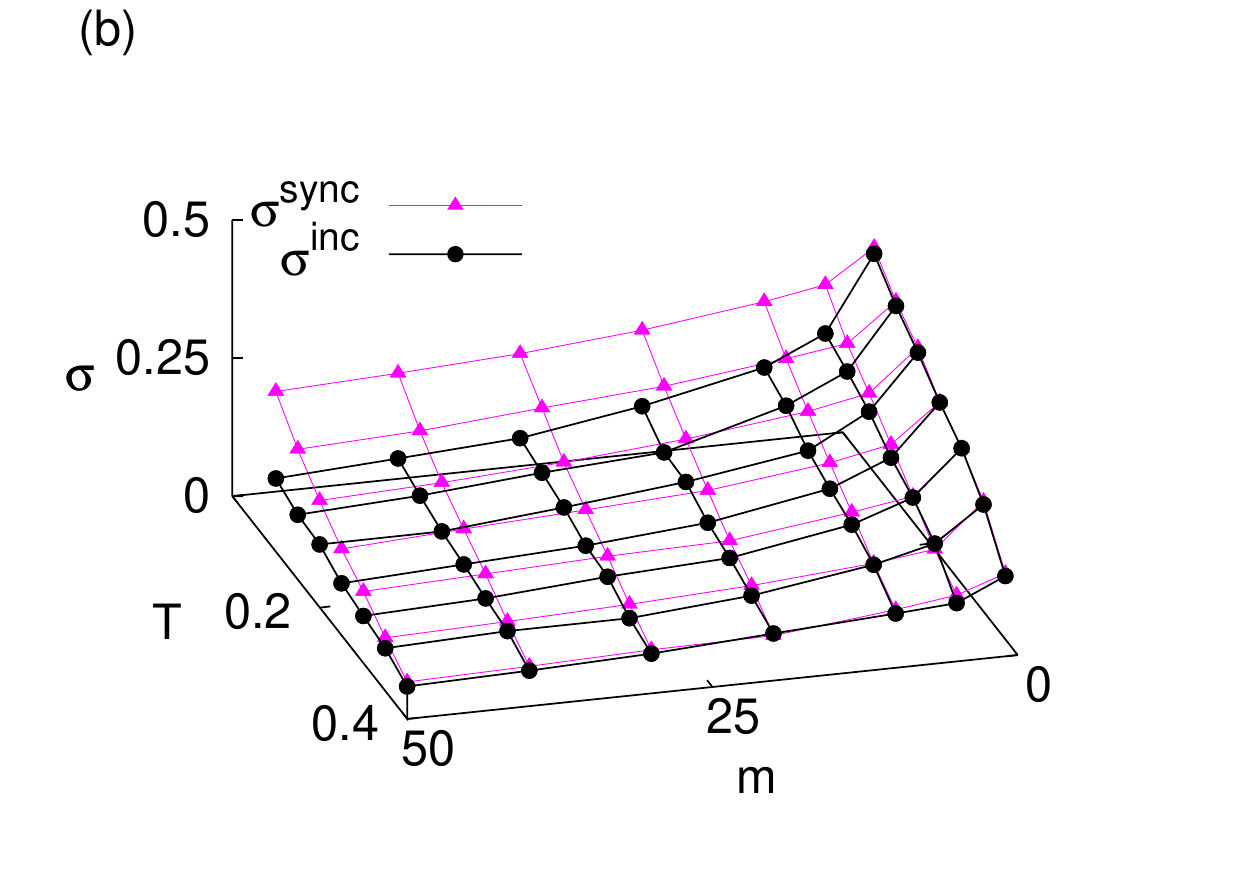}
\caption{Panel (a) shows the schematic phase diagram of model
(\ref{eq:eom-scaled}) in the three-dimensional space of the parameters,
the dimensionless moment of inertia $m$, the temperature $T$, and the width of the
frequency distribution $\sigma$. The shaded blue
surface is a first-order transition surface, and the thick red lines are
second-order critical lines. The system is synchronized inside the region
bounded by the surface, and is incoherent outside. The figure also shows
the transitions of
known models discussed in the text. The blue surface in (a) is
bounded from above and below by the dynamical stability thresholds
$\sigma^{\rm sync}(m,T)$ and $\sigma^{\rm inc}(m,T)$ of respectively the
synchronized and the incoherent phase, which are estimated in $N$-body
simulations from hysteresis plots (see Fig. \ref{fig:hys-mTvary} for an example). The
surfaces $\sigma^{\rm sync}(m,T)$ and $\sigma^{\rm inc}(m,T)$ obtained
in numerical simulations for
$N=500$ and with a Gaussian $g(\omega)$ with zero mean and unit
width are shown in panel (b). \url{https://doi.org/10.1088/1742-5468/14/08/R08001} {\it \textcopyright  SISSA Medialab Srl.  Reproduced
by permission of IOP Publishing.  All rights reserved.}}
\label{fig:phdiag}
\end{figure}

\begin{figure}[!ht]
\centering
\includegraphics[width=90mm]{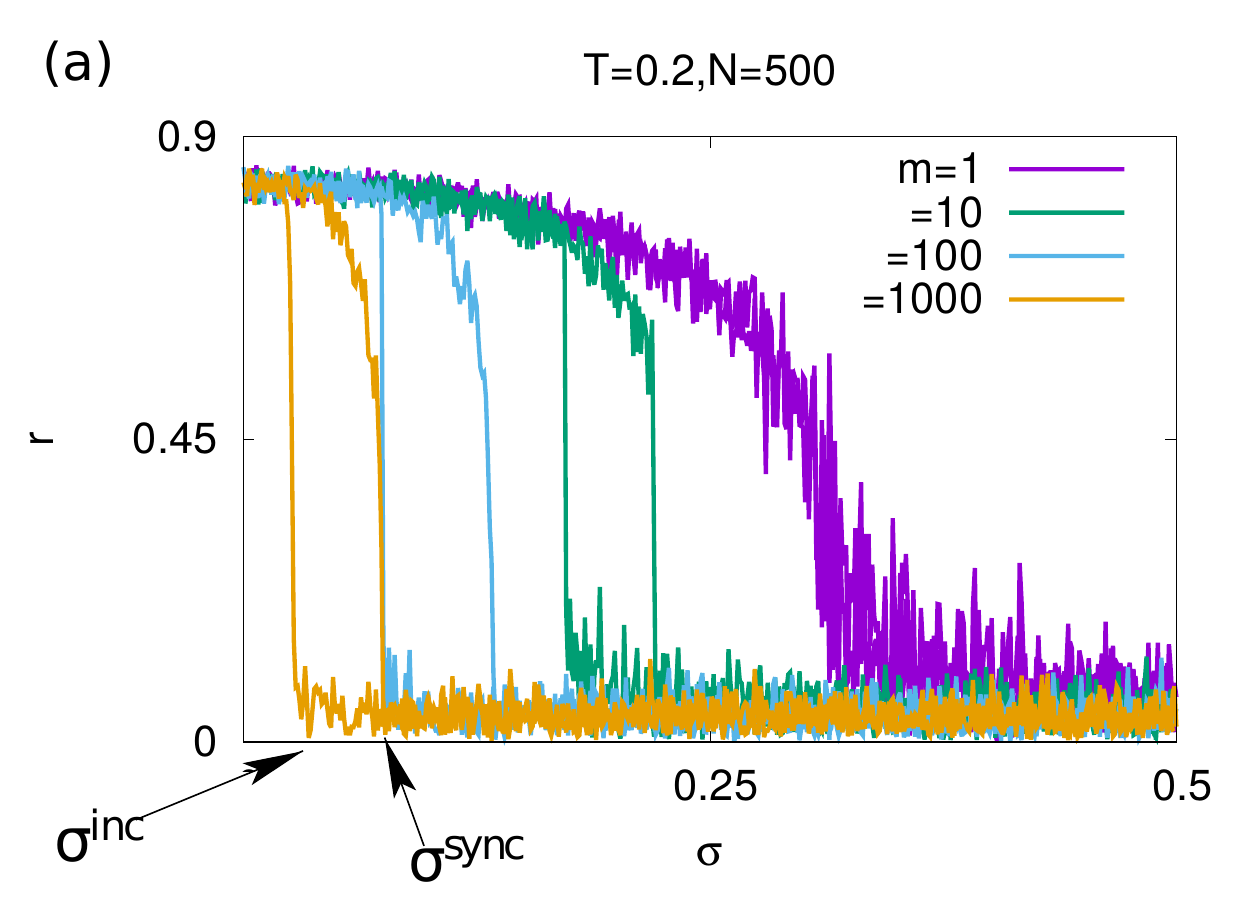}
\includegraphics[width=90mm]{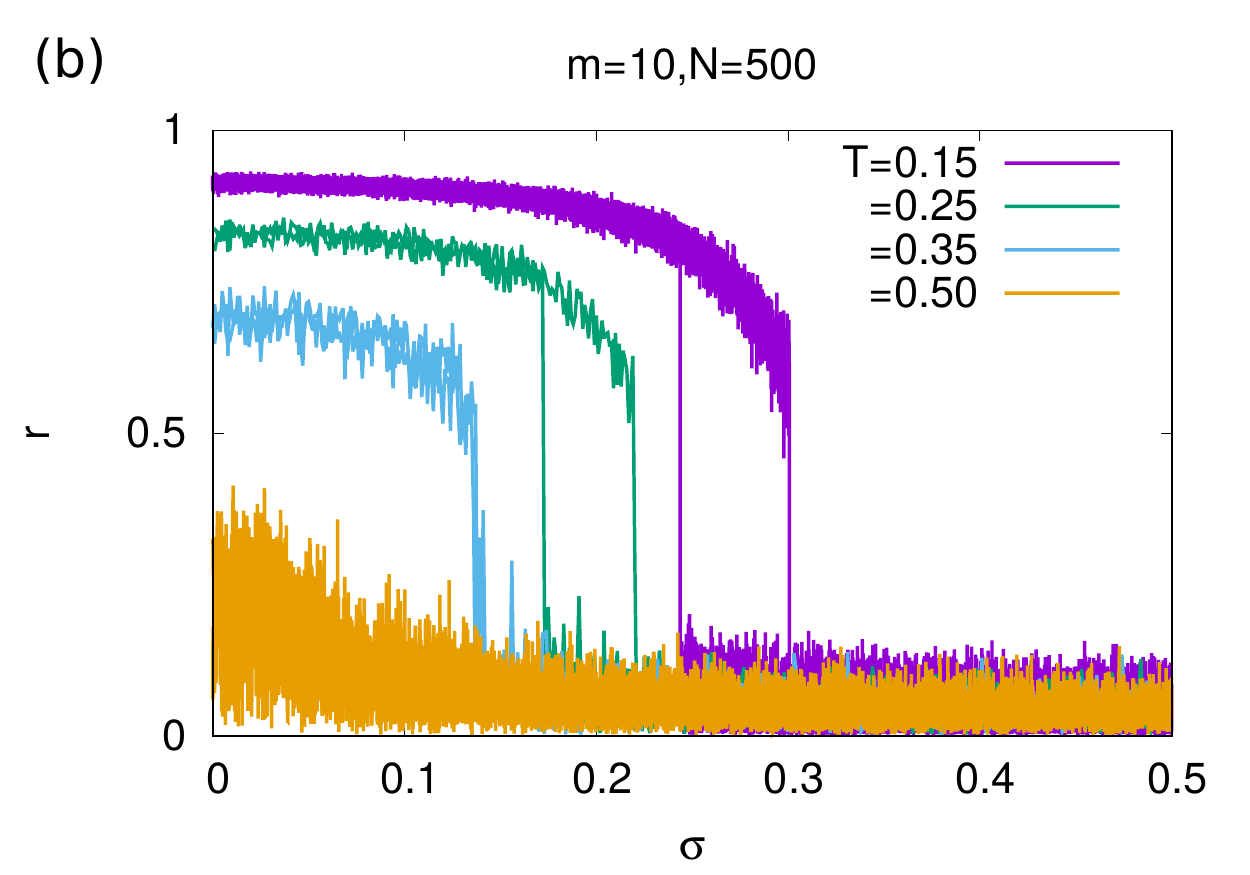}
\caption{For the model (\ref{eq:eom-scaled}), the figure shows (a) $r$ vs.
adiabatically-tuned $\sigma$ for different values of $m$ at
$T=0.2<T_c=1/2$ (with $T_c$ being the BMF transition point), and also the stability thresholds, $\sigma^{\rm inc}(m,T)$ and
$\sigma^{\rm sync}(m,T)$, for $m=1000$, and (b) $r$ vs. adiabatically tuned
$\sigma$ for different temperatures $T \le T_c=1/2$ at a fixed moment of
inertia $m=10$. For a given $m$ in (a), the branch of the plot to
the right (left) corresponds to $\sigma$ increasing (decreasing); for
$m=1$, the two branches almost overlap. For a given $T$ in (b), the branch of the plot to
the right (left) corresponds to $\sigma$ increasing (decreasing); for $T
\ge 0.45$, the two branches practically overlap. The data are obtained from
numerical integration of the dynamics~(\ref{eq:eom-scaled}) for $N=500$
and a Gaussian $g(\omega)$ with zero mean and unit width. \url{https://doi.org/10.1088/1742-5468/14/08/R08001} {\it \textcopyright  SISSA Medialab Srl.  Reproduced
by permission of IOP Publishing.  All rights reserved.}}
\label{fig:hys-mTvary}
\end{figure}

\begin{figure}[!ht]
\centering
\includegraphics[width=90mm]{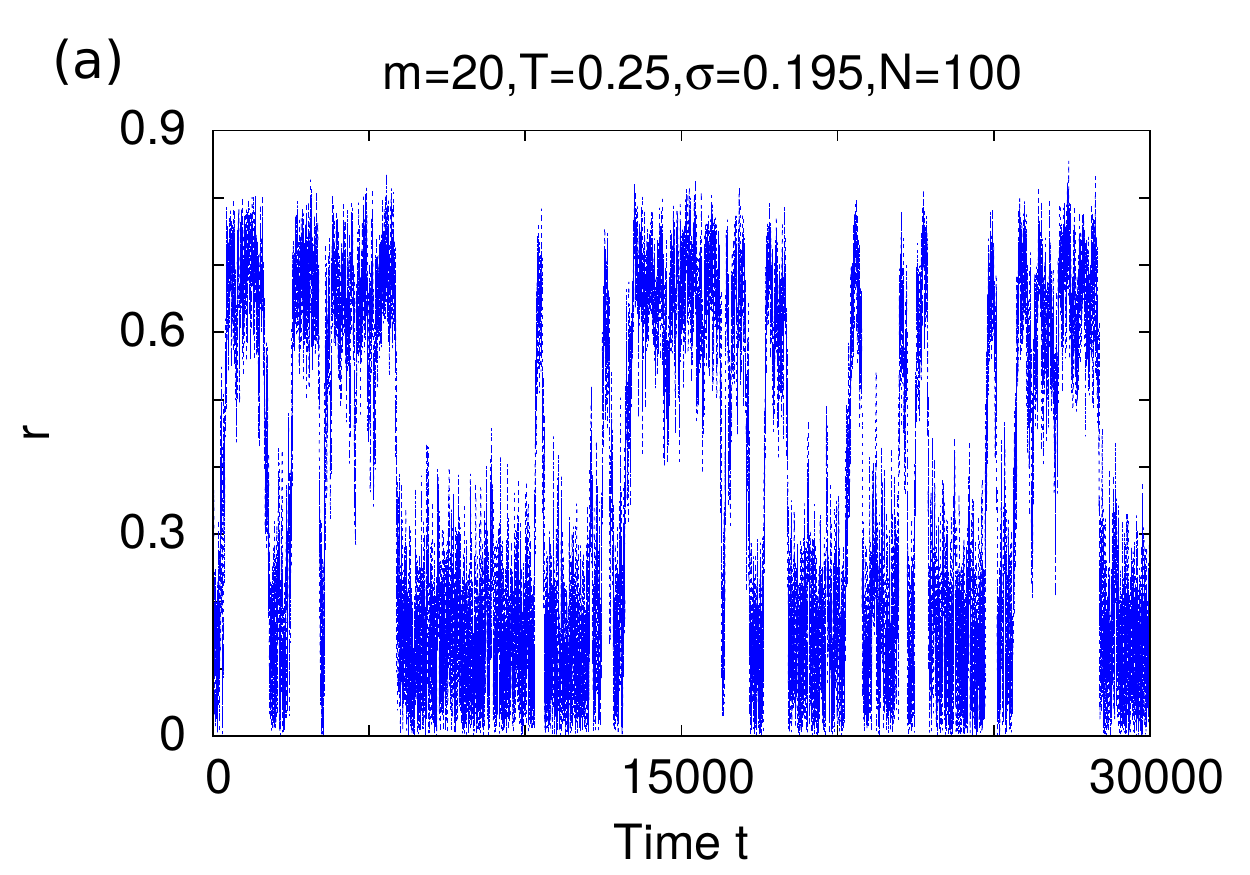}
\includegraphics[width=90mm]{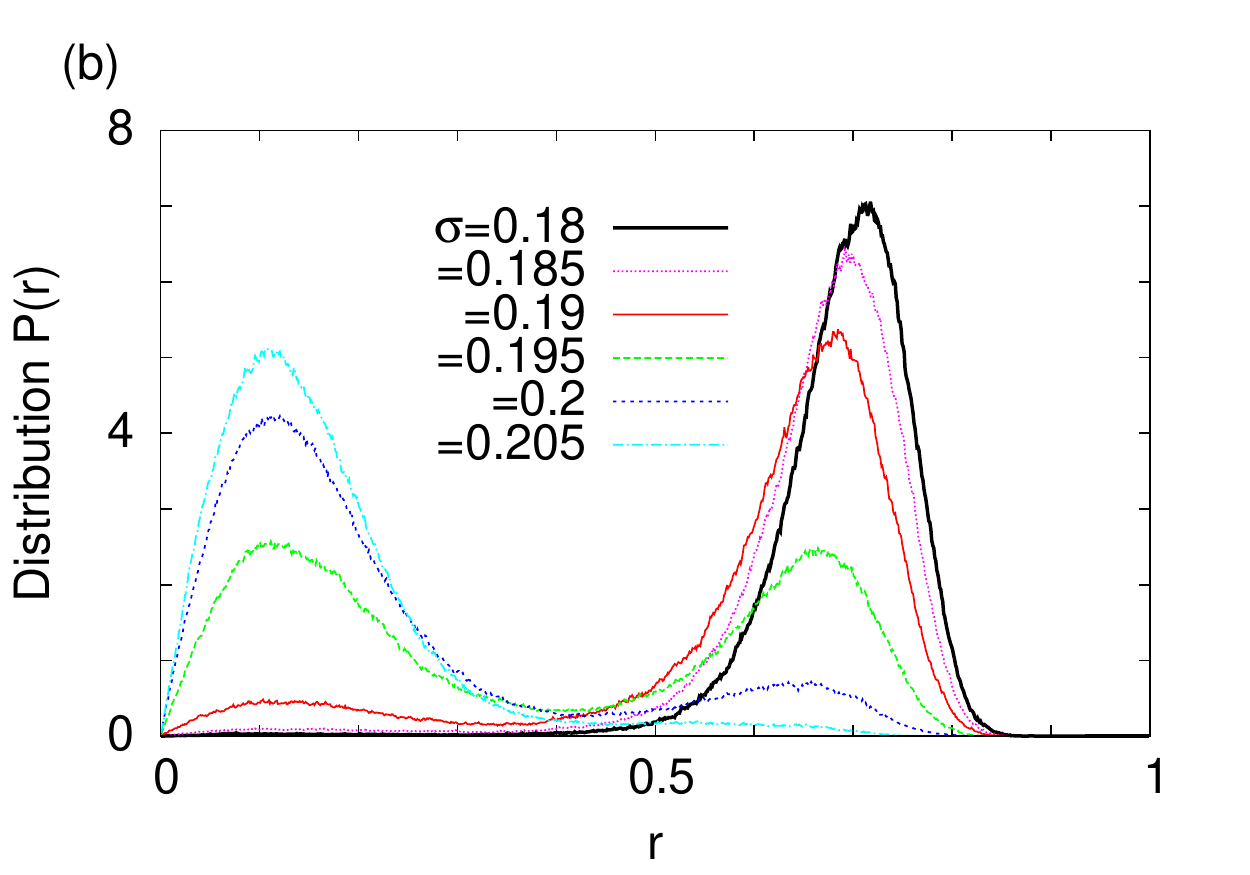}
\caption{For the dynamics (\ref{eq:eom-scaled}) at $m=20,T=0.25,N=100$,
and for a Gaussian $g(\omega)$ with zero mean and unit width, panel (a) shows at 
$\sigma=0.195$, which is the numerically estimated first-order phase transition
point, the quantity $r$ as a function of time in the stationary state, while panel (b)
shows the distribution $P(r)$ at several $\sigma$'s 
around $0.195$. The data are obtained from numerical
integration of the dynamical equations~(\ref{eq:eom-scaled}) with
$N=100$. \url{https://doi.org/10.1088/1742-5468/14/08/R08001} {\it \textcopyright  SISSA Medialab Srl.  Reproduced
by permission of IOP Publishing.  All rights reserved.}}
\label{fig:r-vs-t-pr}
\end{figure}

Figure~\ref{fig:phdiag}(a) shows the complete phase diagram of the
model~(\ref{eq:eom-scaled}), in which the thick red second-order critical lines denote the continuous
transitions mentioned above~\cite{Gupta:2014,Gupta:2014-2}. For $m,\sigma,T$ all
non-zero, however, the synchronization transition becomes first order,
occurring across the shaded blue transition surface. The surface is bounded by the
second-order critical lines on the $(T,\sigma)$ and $(m,T)$ planes, and by a first-order transition line on
the $(m,\sigma)$-plane. Let us remark that all phase
transitions for $\sigma \ne 0$ are in NESSs.

The first-order nature of the phase transition becomes evident on
analyzing results of $N$-body simulations of the
dynamics~(\ref{eq:eom-scaled}) for a representative $g(\omega)$, for
example, a Gaussian distribution
$g(\omega)=\exp(-\omega^2/2)/\sqrt{2\pi}$~\cite{Gupta:2014,Gupta:2014-2}. For given values of $m$ and
$T$, an initial state, which has all the oscillators at $\theta=0$
and angular velocities $v_i$'s sampled from a Gaussian distribution with zero mean and
standard deviation $\propto T$, was first allowed to equilibrate at
$\sigma=0$. The state was subsequently allowed to evolve under the
condition of $\sigma$ increasing adiabatically to high values and back
in a cycle. In Fig.~\ref{fig:hys-mTvary}(a), we show the behavior of $r$ for several $m$'s at a fixed value of $T$ smaller than the BMF transition point $T_c=1/2$. In the figure, one may observe sharp jumps and
hysteresis behavior reminiscent of a first-order transition. With
decrease of $m$, one may observe that the jumps in $r$ become less
sharp, and the hysteresis loop area decreases, both features being
consistent with the fact that the transition becomes second-order-like
as $m \to 0$, see Fig.~\ref{fig:phdiag}(a). For $m=1000$, we show in
Fig.~\ref{fig:hys-mTvary}(a) the approximate stability thresholds for the 
incoherent and the synchronized state, which are denoted respectively by $\sigma^{\rm inc}(m,T)$ and
 $\sigma^{\rm sync}(m,T)$. The actual phase
transition point $\sigma_c(m,T)$ lies in between the two thresholds. Let
us note from the figure that both the thresholds decrease and approach zero with
the increase of $m$. Figure~\ref{fig:hys-mTvary}(b) shows hysteresis plots for a Gaussian $g(\omega)$ at a
fixed $m$ and for several values of $T \le T_c$: one observes
that with $T$ approaching $T_c$, the hysteresis loop area decreases, jumps in $r$ become less sharp and
occur between smaller and smaller values that approach zero. Moreover, the
$r$ value at $\sigma=0$ decreases as $T$ increases towards $T_c$,
reaching zero at $T_c$. These findings imply that the thresholds $\sigma^{\rm inc}(m,T)$ and  $\sigma^{\rm sync}(m,T)$ coincide on the second-order critical lines, as  expected, and moreover, they come asymptotically close together and approach zero in the limit $m \to \infty$ at a fixed $T$. 
For given values of $m$ and $T$ and $\sigma$ in the range
$\sigma^{\rm inc}(m,T) < \sigma < \sigma^{\rm sync}(m,T)$, we show in
Fig.~\ref{fig:r-vs-t-pr}(a) the quantity $r$ as a function of time in the
stationary state. One may observe from the figure a  bistable behavior,
with the system switching back and forth between incoherent ($r \approx 0$)
and synchronized ($r > 0$) states.  Consistently, the distribution $P(r)$ shown in
Figure~\ref{fig:r-vs-t-pr}(b) is indeed bimodal with a peak around
either $r \approx 0$ or $r>0$ as $\sigma$ varies between $\sigma^{\rm
inc}(m,T)$ and $\sigma^{\rm sync}(m,T)$. Figure \ref{fig:r-vs-t-pr} lends
further evidence in support of the phase transition being first
order~\cite{Goldenfeld:1992}.

\subsection{Analysis in the continuum limit: The Kramers equation}
\label{sec:analysis}
In this section, we discuss analytical characterization of the
dynamics~(\ref{eq:eom-scaled}) in the continuum limit $N \to \infty$.
Similar to what was done for the Kuramoto model, we define a
single-oscillator density $f(\theta,v,\omega,t)$ that 
gives at time $t$ and for each $\omega$ the fraction of oscillators that
have angle $\theta$
and angular velocity $v$. The density $f$ is $2\pi$-periodic in
$\theta$, obeys the normalization $\int_{-\pi}^{\pi} {\rm d}\theta
\int_{-\infty}^{+\infty} {\rm d}
v~f(\theta,v,\omega,t)=1~\forall~\omega,t$, and has a time
evolution given by the so-called Kramers equation~\cite{Risken:1996,Acebron:2000,Gupta:2014,Gupta:2014-2}
\begin{equation}
\frac{\partial f}{\partial t}=-v\frac{\partial f}{\partial
\theta}+\frac{\partial}{\partial
v}\Big(\frac{v}{\sqrt{m}}-\sigma \omega+r\sin(\theta-\psi)\Big)f+\frac{
T}{\sqrt{m}}\frac{\partial^2 f}{\partial v^2}\,, 
\label{eq:Kramers}
\end{equation}
with $r(t)e^{i\psi(t)}=\int {\rm d}\theta {\rm d} v {\rm
d}\omega~g(\omega)e^{i\theta}f(\theta,v,\omega,t)$.

We are interested in the stationary state solutions of the Kramers
equation, obtained by setting the left hand side of
Eq.~(\ref{eq:Kramers}) to zero. As already mentioned, the
stationary state is a NESS, unless $\sigma=0$. In the stationary state, the
quantities $r$ and $\psi$ have their stationary-state values $r_{\rm
st}$ and $\psi_{\rm st}$, respectively. The stationary-state single-oscillator
density $f_{\rm st}(\theta,v,\omega)$ thus satisfies
\begin{equation}
0=-v\frac{\partial f_{\rm st}}{\partial
\theta}+\frac{\partial}{\partial
v}\Big(\frac{v}{\sqrt{m}}-\sigma \omega+r_{\rm st}\sin (\theta-\psi_{\rm
st})
\Big)f_{\rm st}+\frac{
T}{\sqrt{m}}\frac{\partial^2 f_{\rm st}}{\partial v^2}\,. 
\label{eq:Kramers-stationary}
\end{equation}
Similar to what was done in
Section~\ref{sec:thermodynamic-limit-kuramoto}, we may set $\psi_{\rm st}$ to
zero by choosing suitably the origin of the angle axis, which corresponds to having the stationary values $r_{y,{\rm st}}=0$ and
$r_{x,{\rm st}}=r_{\rm st}$, see Eq.~(\ref{eq:rxry}). Consequently,
one has
\begin{equation}
r_{\rm st}=\int {\rm d}\theta {\rm d} v {\rm
d}\omega~g(\omega)\cos \theta f_{\rm st}(\theta,v,\omega)\,.
\label{eq:rst-stationary}
\end{equation}
From now on, we will consider the stationary-state Kramers equation with
$\psi_{\rm st}=0$.

\subsection{$\sigma=0$: Stationary solutions and the associated phase
transition}
\label{sec:BMF-stationary-state}
For $\sigma=0$, the stationary-state single-oscillator density is given by
the Gibbs-Boltzmann measure corresponding to canonical
equilibrium~\cite{Gupta:2014-2}:
\begin{equation}
f_{\rm st}(\theta,v)=\frac{\exp [-v^2/(2T)+(r_{\rm st}/T)\cos
\theta]}{\sqrt{2\pi T}\int_{-\pi}^\pi {\rm d}\theta~\exp[(r_{\rm
st}/T)\cos \theta]}\,,
\label{eq:sigma0-stationary-again}
\end{equation}
where the denominator is the normalization factor that
ensures that $\int_{-\infty}^\infty {\rm d}v \int_{-\pi}^\pi {\rm
d}\theta~f_{\rm st}(\theta,v)=1$. 
One may easily check by direct substitution that the above form\footnote{Note that with $\sigma=0$, all the oscillators have the same
natural frequency equal to $\langle \omega \rangle$, and the need to
group the oscillators based on their natural frequencies, as was done
for defining the density $f(\theta,v,\omega,t)$, is no longer there.
Consequently, one has the stationary-state single-oscillator density
denoted by $f_{\rm st}(\theta,v)$ and which is defined as the fraction of oscillators
that have angle $\theta$ and angular velocity $v$ in the stationary
state.} satisfies Eq.~(\ref{eq:Kramers-stationary}) with $\sigma=0$ and
with $\psi_{\rm st}=0$. Using
Eqs.~(\ref{eq:rst-stationary}) and~(\ref{eq:sigma0-stationary-again}), we get 
\begin{equation}
r_{\rm st}=\int {\rm d}\theta {\rm d} v~\cos\theta f_{\rm
st}(\theta,v)=\frac{\int_{-\pi}^\pi {\rm d}\theta~\cos \theta
\exp[(r_{\rm st}/T)\cos \theta]}{\int_{-\pi}^\pi {\rm
d}\theta~\exp[(r_{\rm st}/T) \cos
\theta]}\,.
\label{eq:rst-self-consistency}
\end{equation}
The self-consistency condition, Eq.~(\ref{eq:rst-self-consistency}), has
a trivial solution $r_{\rm st}=0$ valid at all temperatures, while it
may be shown that a non-zero solution exists for $T$ smaller than a
critical value $T_c=1/2$~\cite{Campa:2009}. Reverting to dimensional
temperatures by using Eq.~(\ref{eq:dimensionless-definition}), we obtain
the critical temperature of the BMF model as 
$T_c=\widetilde{K}/2$, as announced towards the end of
Section~\ref{sec:longrangemodel}.

\subsection{$\sigma \ne 0$: Incoherent stationary state and its linear stability}
\label{sec:incoherent-stationary-state}
For $\sigma \ne 0$, the $\theta$-independent solution characterizing the incoherent phase, for which
$r_{\rm st}=0$, is given by~\cite{Acebron:2000}:
\begin{equation}
f^{\rm inc}_{\rm st}(\theta,v,\omega)=\frac{1}{2\pi}\sqrt{\frac{1}{2\pi T}}
\exp \left[-\frac{(v-\sigma \omega \sqrt{m})^2}{2T}\right]\,.
\label{eq:inc-st}
\end{equation}
The linear stability analysis of the incoherent state~(\ref{eq:inc-st}) may be carried out by expanding
$f(\theta,v,\omega,t)$ as $f(\theta,v,\omega,t)=f^{\rm inc}_{\rm
st}(\theta,v,\omega)+e^{\lambda
t}\delta f(\theta,v,\omega)$, with $|\delta f| \ll 1$, substituting in
Eq.~(\ref{eq:Kramers}), and keeping terms to linear order in $\delta f$. The solution of
the linearized equation yields the following equation that $\lambda$ has
to satisfy~\cite{Acebron:2000}: 
\begin{equation}
\frac{2T}{e^{mT}}=\sum_{p=0}^\infty
\frac{(-m T)^p(1+\frac{p}{mT})}{p!}\int\limits_{-\infty}^{\infty}
\frac{g(\omega){\rm d}\omega}{
1+\frac{p}{mT}+i\frac{\sigma\omega}{T}+\frac{\lambda}{T\sqrt{m}}}\,.
\label{eq:stability-eqn}
\end{equation}

A rather long analysis allows one to prove that the above equation has one
and only one solution for $\lambda$ with a positive real part, and when
this single solution exists, it is necessarily real~\cite{Gupta:2014,Gupta:2014-2}. A positive
(respectively, negative) $\lambda$ implies that the incoherent
state~(\ref{eq:inc-st}) is linearly unstable (respectively, stable). It
then follows that at the point of neutral 
stability, one has $\lambda=0$, which when substituted in
Eq.~(\ref{eq:stability-eqn}) gives $\sigma^{\rm inc}(m,T)$, the stability
threshold of the incoherent stationary state, satisfying 
\begin{equation}
\frac{2T}{e^{mT}}=\sum_{p=0}^\infty
\frac{(-mT)^p(1+\frac{p}{mT})^2}{p!}\int\limits_{-\infty}^{\infty}
\frac{g(\omega){\rm d}\omega}{(1+\frac{p}{mT})^2+\frac{(\sigma^{\rm
inc})^2\omega^2}{T^2}}\,.
\label{eq:sigma_inc}
\end{equation}
In the $(m,T,\sigma)$ space, the above equation defines the stability surface $\sigma^{\rm
inc}(m,T)$. There will similarly be the stability surface $\sigma^{\rm
sync}(m,T)$ representing the stability threshold of the synchronized
stationary state. The reader may refer to Fig.~\ref{fig:phdiag}(b) that shows the two surfaces obtained in $N$-body simulations for $N=500$ for a Gaussian $g(\omega)$. 

The two surfaces, $\sigma^{\rm
inc}(m,T)$ and $\sigma^{\rm
sync}(m,T)$, coincide on the critical lines on the
$(T,\sigma)$ and $(m,T)$ planes where the transition becomes continuous,
while outside these planes, the surfaces enclose the first-order transition surface
$\sigma_c(m,T)$, that is, $\sigma^{\rm sync}(m,T) > \sigma_c(m,T) >\sigma^{\rm
inc}(m,T)$, see Fig.~\ref{fig:phdiag}(a). In this regard, let us show by taking suitable
limits that the surface $\sigma^{\rm
inc}(m,T)$ meets the critical lines on the $(T,\sigma)$ and
$(m,T)$ planes. We will also obtain the intersection of this surface with the $(m,\sigma)$-plane.
On considering $m \to 0$ at a fixed $T$, noting that only the $p=0$ term in the
sum in Eq.~(\ref{eq:sigma_inc}) contributes yields $\lim_{m \to 0, T
\,{\rm fixed}}\sigma^{\rm inc}(m,T)=\sigma_c(m=0,T)$, with the implicit
expression of $\sigma_c(m=0,T)$ given by Eq.~(\ref{eq:sakaguchi-again}). One
also finds that $\displaystyle \lim_{T \to T_c^-, m \,{\rm
fixed}}\sigma^{\rm inc}(m,T)=0$, that is, on the $(m,T)$ plane, the transition line
is given by $T_c=1/2$. When $T \to 0$ at a fixed $m$, we get
$\displaystyle \sigma^{\rm inc}_{\rm noiseless}(m)\equiv \lim_{T \to 0, m \,{\rm fixed}}\sigma^{\rm
inc}(m,T)$, with~\cite{Gupta:2014,Gupta:2014-2}.
\begin{equation}
1=\frac{\pi g(0)}{2\sigma^{\rm inc}_{\rm
noiseless}}-\frac{m}{2}\int_{-\infty}^{\infty}{\rm
d}\omega~\frac{g(\omega)}{\left[1+m^2(\sigma^{\rm
inc}_{\rm noiseless})^2\omega^2\right]}\,.
\end{equation}

\subsection{$\sigma \ne 0$: Synchronized stationary state}
\label{sec:synchronized-stationary-state}
For $\sigma \ne 0$, the existence of the synchronized stationary state is borne out by our
simulation results shown in Figs.~\ref{fig:hys-mTvary}
and~\ref{fig:r-vs-t-pr}. For general $\sigma$,
we expand the single-oscillator density for the synchronized stationary
state as~\cite{Campa:2015} 
\begin{equation}
f^{\rm sync}_{\rm st}(\theta,v,\omega)=
\Phi_0\left( \frac{v}{\sqrt{2T}}\right) \sum_{n=0}^\infty b_n(\theta,\omega)
\Phi_n\left( \frac{v}{\sqrt{2T}}\right)\,.
\label{eq:f-expansion}
\end{equation}
Here, the functions $b_n$ satisfy
$b_n(\theta,\omega)=b_n(\theta+2\pi,\omega)$ to ensure that $f^{\rm
sync}_{\rm st}$ is
$2\pi$-periodic in $\theta$, while $\Phi_n(ax)$ is the Hermite function:
$\Phi_n(ax)=\sqrt{a/(2^n n!\sqrt{\pi})}\exp\left[-\frac{a^2x^2}{2}\right]H_n(ax)$, with
$H_n(x)$'s being the $n$-th degree Hermite polynomial. The functions $\Phi_n$ are orthonormal: $\int {\rm d} x ~\Phi_m(ax) \Phi_n(ax) = \delta_{mn}$.
Normalization of $f^{\rm sync}_{\rm st}(\theta,v,\omega)$ implies the equality
$\int_{-\pi}^\pi {\rm d} \theta ~b_0(\theta,\omega)=1$, while the
self-consistent values of the parameters $r_{\rm st}$ are given by
\begin{equation}
r_{\rm st}=\int {\rm d} \omega~g(\omega)\int_{-\pi}^\pi{\rm d}
\theta~b_0(\theta,\omega) \cos \theta\,.
\label{eq:rstself}
\end{equation}
Furthermore, using $\int {\rm d} x~x\Phi_0(ax)\Phi_n(ax)=1/(\sqrt{2}a)\delta_{n,1}$, 
we obtain that $\int {\rm d} v~vf^{\rm sync}_{\rm st}(\theta,v,\omega) =
\sqrt{T}b_1(\theta,\omega)$. On the other hand, integrating over $v$ the
stationary-state Kramers equation~(\ref{eq:Kramers-stationary}), we obtain that
$\int {\rm d} v~vf^{\rm sync}_{\rm st}(\theta,v,\omega)$ and, hence, $b_1(\theta,\omega)$, does not depend on $\theta$.
Choosing the Hermite functions in the expansion~(\ref{eq:f-expansion})
is motivated by the fact that for $\sigma=0$, the
density $f_{\rm st}^{\rm sync}(\theta,v,\omega)$ has the Gibbs-Boltzmann
form, $f_{\rm st}^{\rm sync}(\theta,v,\omega) \sim \exp
[-v^2/(2T)+r_{\rm st}\cos \theta]$, cf.
Eq.~(\ref{eq:sigma0-stationary-again}). As
may be shown~\cite{Campa:2015}, the expansion coefficients $b_n$ for this case satisfy
$b_0(\theta,0) \sim \exp [r_{\rm st} \cos \theta], ~b_n(\theta,0)=0$ for $n>0$, so that only
the $n=0$ term in the expansion (\ref{eq:f-expansion}) has to be taken into
account; then, with $\Phi_0(x)\sim\exp(-x^2/2)$, the product
$\Phi_0\left( v/\sqrt{2T}\right)\Phi_0\left(v/\sqrt{2T}\right)$
appearing in the expansion correctly
reproduces the velocity-part of the density $\sim \exp [-v^2/(2T)]$.

On plugging the expansion (\ref{eq:f-expansion}) into the stationary-state Kramers
equation~(\ref{eq:Kramers-stationary}), on using the known recursion relations for the Hermite
polynomials, and on equating to zero the coefficient of each $\Phi_n$, we
get~\cite{Campa:2015}
\begin{eqnarray}
\sqrt{nT}\frac{\partial b_{n-1}(\theta,\omega)}{\partial \theta}
+\sqrt{(n+1)T}\frac{\partial b_{n+1}(\theta,\omega)}{\partial
\theta}+\frac{n}{\sqrt{m}}b_n(\theta,\omega)
+\sqrt{\frac{n}{T}}b_{n-1}(\theta,\omega)[r_{\rm st}\sin \theta-\sigma \omega]=0 \nonumber \\
\label{eq:bp-eqn}
\end{eqnarray}
for $n=0,1,2,\dots$ (with the understanding that $b_{-1}(\theta,\omega)\equiv 0$). The equation for $n=0$ recovers the result that
$b_1(\theta,\omega)$ is independent of $\theta$. Noting the scaling of
the various terms in Eq.~(\ref{eq:bp-eqn}) with $m$, we expand
$b_n(\theta,\omega)$
as~\cite{Campa:2015}
\begin{equation}
b_n(\theta,\omega)=\sum_{k=0}^\infty (\sqrt{m})^k
c_{n,k}(\theta,\omega)\,,
\label{eq:bexpansion}
\end{equation}
which may be shown to be an asymptotic expansion in
$\sqrt{m}$~\cite{Campa:2015}, thus
requiring a proper numerical evaluation of the sum on the rhs by invoking the so-called Borel
summation method~\cite{Hardy:1991}. Now, using
Eq.~(\ref{eq:bexpansion}), we conclude that $b_1(\theta,\omega)$ being
independent of $\theta$ implies that so
is $c_{1,k}(\theta,\omega)~\forall~k$.
The only constraint on $b_0(\theta,\omega)$ being $\int_{-\pi}^{\pi}
{\rm d} \theta ~b_0(\theta,\omega)=1$, we may
without loss of generality choose $c_{0,k \ge 1}(0,\omega)=0$. We now
use Eq.~(\ref{eq:bexpansion}) in Eq.~(\ref{eq:bp-eqn}) and equate to zero the coefficient
of each power of $\sqrt{m}$. The term proportional to
$\left(\sqrt{m}\right)^{-1}$ gives simply $nc_{n,0}(\theta,\omega)=0$,
which implies that we have $c_{n,0}(\theta,\omega)=0$ for $n>0$. The coefficient of the
term proportional to $\left(\sqrt{m}\right)^k$ leads
to~\cite{Campa:2015}
\begin{eqnarray}
\sqrt{nT}\frac{\partial c_{n-1,k}(\theta,\omega)}{\partial \theta}
+\sqrt{(n+1)T}\frac{\partial c_{n+1,k}(\theta,\omega)}{\partial \theta}
+\sqrt{nT} a(\theta,\omega) c_{n-1,k}(\theta,\omega) + nc_{n,k+1}(\theta,\omega)=0 \nonumber \\
\label{eq:finalsystem}
\end{eqnarray}
for $n,k=0,1,2,\ldots$ (with $c_{-1,k}(\theta,\omega)\equiv 0$),
where $a(\theta,\omega)\equiv[r_{\rm st} \sin \theta-\sigma \omega]/T$. The system of
equations (\ref{eq:finalsystem})
can be solved recursively. While the details of solving these equations
may be found in Ref.~\cite{Campa:2015}, we quote here only the
solutions:
\begin{eqnarray}
c_{0,0}(\theta,\omega)&=&c_{0,0}(0,\omega)e^{-
h(\theta,\omega)}\left[1+\left(e^{h(2\pi,\omega)}-1\right)
\frac{\int_0^\theta {\rm d} \theta'
e^{h(\theta',\omega)}}{\int_{-\pi}^{\pi} {\rm d} \theta'
e^{h(\theta',\omega)}}\right]\,, \\
c_{1,1}(\omega)&=&\sqrt{T}\frac{c_{0,0}(0,\omega)\left(1-e^{h(2\pi,\omega)}\right)}
{\int_{-\pi}^\pi {\rm d} \theta' e^{h(\theta',\omega)}}\,, \\
c_{n,n}(\theta,\omega)&=&-\sqrt{\frac{T}{n}}\left[\frac{\partial
c_{n-1,n-1}(\theta,\omega)}{\partial \theta}
+ a(\theta,\omega )c_{n-1,n-1}(\theta,\omega) \right]\,,\\
c_{0,2k}(\theta,\omega)&=&\sqrt{2}\frac{\int_{-\pi}^\pi {\rm d} \theta'
\frac{\partial c_{2,2k}(\theta',\omega)}{\partial \theta'}
e^{h(\theta',\omega)}}{\int_{-\pi}^\pi {\rm d} \theta'
e^{h(\theta',\omega)}}e^{-h(\theta,\omega)}\int_0^\theta {\rm d} \theta'
e^{h(\theta',\omega)} \nonumber \\
&&-\sqrt{2}e^{-h(\theta,\omega)}\int_0^\theta {\rm d} \theta'
\frac{\partial c_{2,2k}(\theta',\omega)}{\partial \theta'}
e^{h(\theta',\omega)}\,, 
\label{eq:solcn02k} \\
c_{1,1+2k}(\omega)&=&-\sqrt{2T}\frac{\int_{-\pi}^\pi {\rm d} \theta'
\frac{\partial c_{2,2k}(\theta',\omega)}{\partial \theta'}
e^{h(\theta',\omega)}}{\int_{-\pi}^\pi {\rm d} \theta'
e^{h(\theta',\omega)}}\,,
\label{eq:solcn11p2k} \\
c_{2,2+2k}(\theta,\omega)&=& -\sqrt{\frac{T}{2}} a(\theta,\omega)
c_{1,1+2k}(\omega)-\frac{\sqrt{3T}}{2}\frac{\partial
c_{3,1+2k}(\theta,\omega)}{\partial \theta}\,,
\label{eq:solcn22p2k} \\ 
c_{n,n+2k}(\theta,\omega)&=&-\sqrt{\frac{T}{n}}\left[ \frac{\partial
c_{n-1,n-1+2k}(\theta)}{\partial \theta}
+a(\theta,\omega)c_{n-1,n-1+2k}(\theta,\omega)\right] \nonumber \\
&&-\frac{\sqrt{(n+1)T}}{n}\frac{\partial c_{n+1,n-1+2k}(\theta,\omega)}{\partial
\theta} \,\,\,\,\,\, n \ge 3\,, 
\label{eq:solcnnp2k}
\end{eqnarray}
with $k=1,2,\dots$. 
Here, we have defined $h(\theta,\omega) \equiv \int_0^\theta {\rm
d}\theta'a(\theta',\omega)$. 

\begin{figure}
\centering
\includegraphics[scale=0.3]{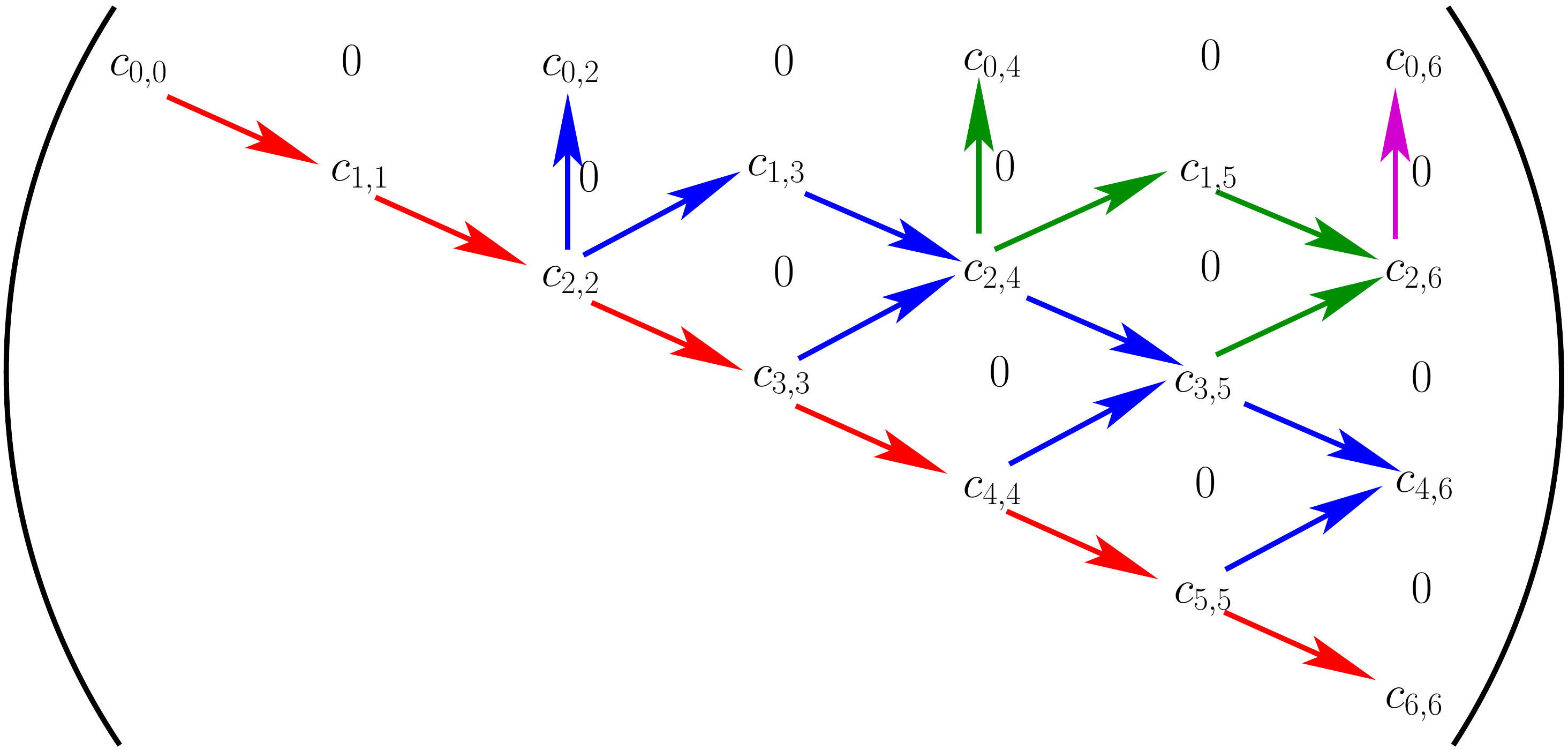}
\caption{Flow diagram for the evaluation of the expansion coefficients
$c_{n,k}(\theta,\omega); n,k=0,1,2,\ldots,6$ by using Eq.~(\ref{eq:finalsystem}). Starting from the main diagonal,
arrows and different colors denote subsequent flows (see text). The
elements below the main diagonal are all zero. \url{
https://doi.org/10.1088/1742-5468/2015/05/P05011} {\it \textcopyright  SISSA Medialab Srl.  Reproduced
by permission of IOP Publishing.  All rights reserved.}}
\label{fig:flowfig}
\end{figure}

Figure~\ref{fig:flowfig} shows schematically the flow of the solution up
to $n=k=6$, while that for higher values proceeds analogously. As shown,
the system~(\ref{eq:finalsystem}) computes progressively each element of the main
diagonal, and then the elements of the second upper diagonal,
each one determined by the knowledge of two previously determined
elements, and so on. Each element of the matrix is proportional to
$c_{0,0}(0,\omega)$, which is fixed by the normalization of
$f_{\rm st}^{\rm sync}$: $\sum_{k=0}^\infty \int_{-\pi}^\pi {\rm d} \theta~(\sqrt{m})^{2k}
c_{0,2k}(\theta,\omega)=1$.
The values of $r_{\rm st}$ have to be determined self-consistently by
using Eqs.~(\ref{eq:rstself}) and (\ref{eq:bexpansion}).

For illustrating an application of the aforementioned scheme, let us choose
a representative $g(\omega)$, namely, a Gaussian:
$g(\omega)=1/(\sqrt{2\pi})\exp(-\omega^2/2)$, and obtain in the
synchronized phase the marginal $\theta$-distribution,
$\displaystyle n(\theta) \equiv \int_{-\infty}^\infty
{\rm d} \omega~g(\omega)\int_{-\infty}^\infty {\rm d} v~f^{\rm
sync}_{\rm st}(\theta,v,\omega)$, and the quantity
$\displaystyle p(\theta) \equiv \int_{-\infty}^\infty
{\rm d} \omega~g(\omega)\int_{-\infty}^\infty {\rm d} v~v^2f^{\rm
sync}_{\rm st}(\theta,v,\omega)$ that is proportional to the local pressure~\cite{Huang:1987}. 
Orthonormality of the Hermite functions
implies that
\begin{eqnarray}
&&n(\theta)=\int_{-\infty}^\infty d\omega~g(\omega)b_0(\theta,\omega)\,, \\
&&p(\theta) = T\int_{-\infty}^\infty d\omega~g(\omega)\left(\sqrt{2}b_2(\theta,\omega)+b_0(\theta,\omega)\right)\,.
\label{eq:exprdens}
\end{eqnarray}
We thus need the coefficients $b_0(\theta,\omega)$ and $b_2(\theta,\omega)$, whose
evaluation requires truncating the expansion~(\ref{eq:bexpansion}) at
suitable values $k_{\rm trunc}$ of $k$. Figure~\ref{fig:flowfig} implies that knowing
$c_{2,2k}$ allows to compute $c_{0,2k}$, so it is natural to choose the same
$k_{\rm trunc}$ for both $b_0(\theta,\omega)$ and $b_2(\theta,\omega)$.

\begin{figure}
\centering
\includegraphics[width=75mm]{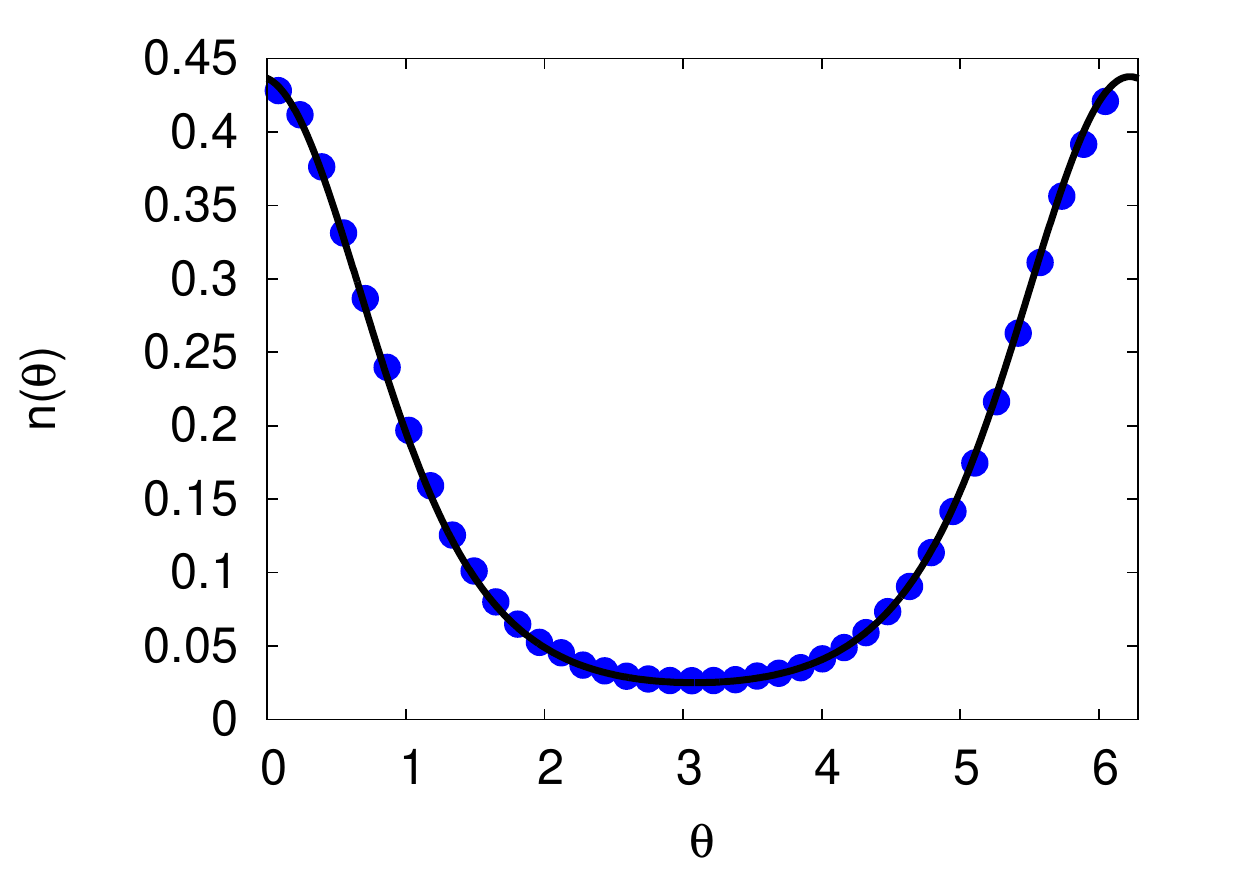}
\includegraphics[width=75mm]{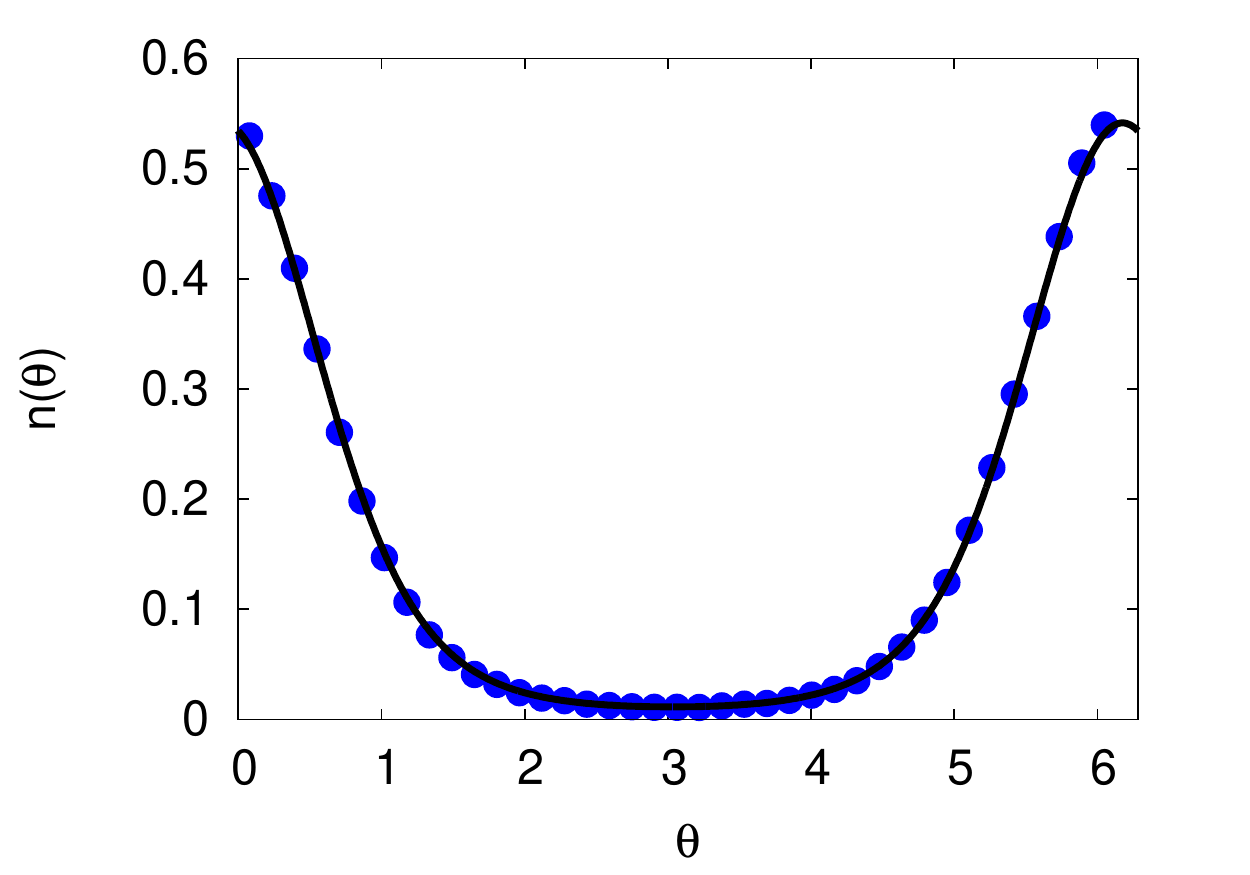}
\caption{Density $n(\theta)$ in the dynamics~(\ref{eq:eom-scaled}) with
a Gaussian $g(\omega)$,
for $m=0.25$, $T=0.25$, $\sigma=0.295$, $k_{\rm trunc}=12$ (left panel), and
for $m=5.0$, $T=0.25$, $\sigma=0.2$, $k_{\rm trunc}=2$ (right panel).
Simulations results are denoted by points and pertain to number of
oscillators $N=10^6$, while theoretical predictions
are denoted by lines. \url{
https://doi.org/10.1088/1742-5468/2015/05/P05011} {\it \textcopyright  SISSA Medialab Srl.  Reproduced
by permission of IOP Publishing.  All rights reserved.}}
\label{fig:fig-ness-th}
\end{figure}

\begin{figure}
\centering
\includegraphics[width=75mm]{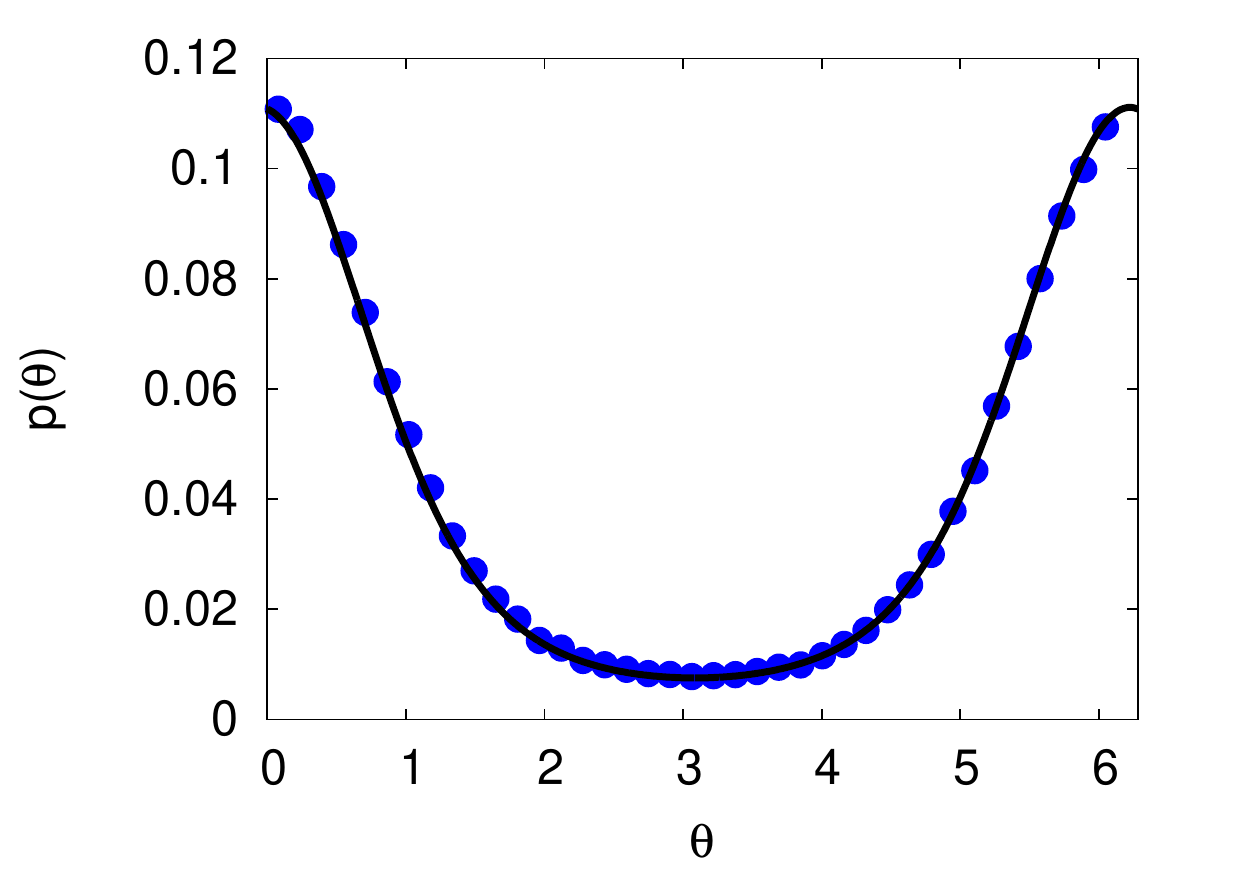}
\includegraphics[width=75mm]{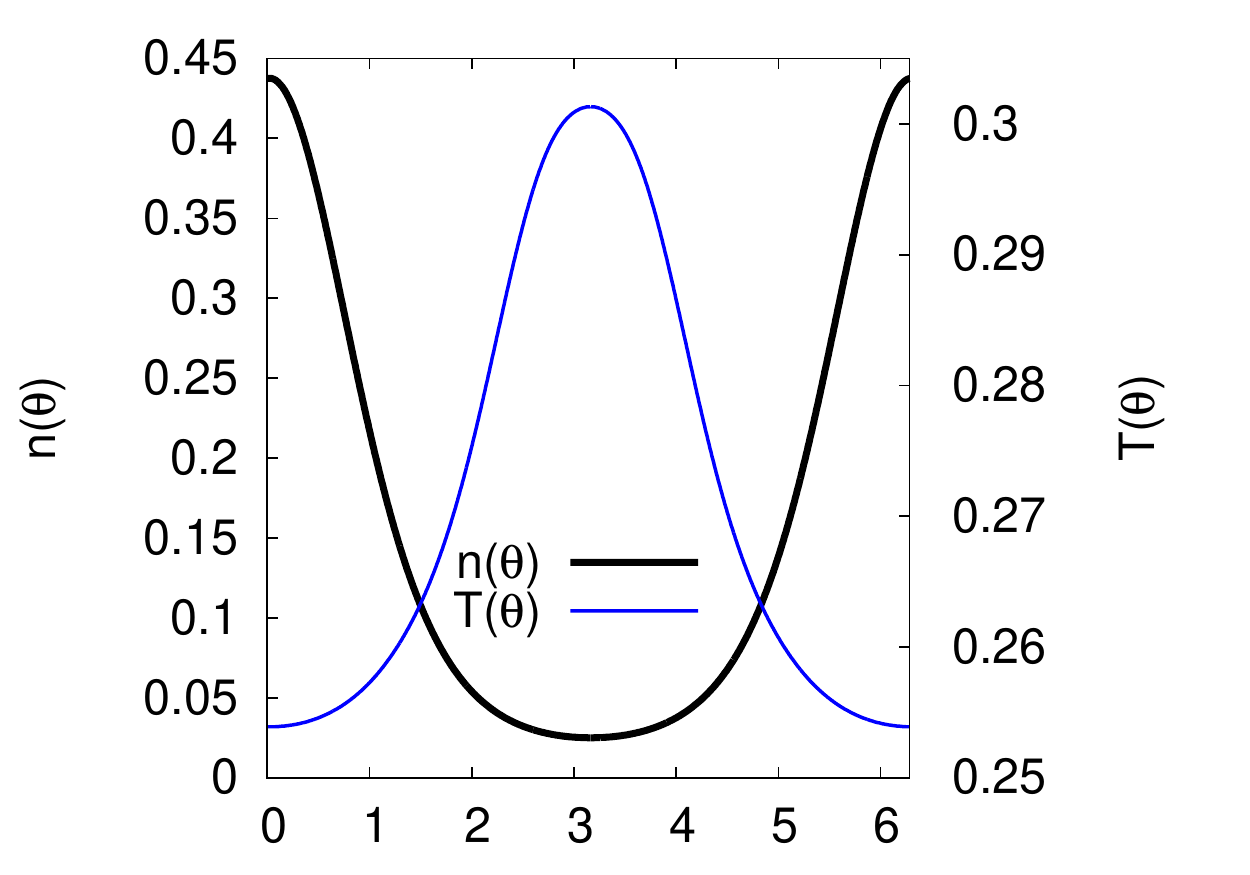}
\caption{In the left panel is shown the pressure $p(\theta)$ for the
same parameters as for the left panel of
Fig.~\ref{fig:fig-ness-th}. Simulation results are depicted by points
and pertain to number of oscillators $N=10^6$, while theoretical predictions
are denoted by lines. In the right panel is shown the local temperature
$T(\theta)=p(\theta)/n(\theta)$ and its
anticorrelation with the density $n(\theta)$. \url{
https://doi.org/10.1088/1742-5468/2015/05/P05011} {\it \textcopyright  SISSA Medialab Srl.  Reproduced
by permission of IOP Publishing.  All rights reserved.}}
\label{fig:fig-ness-temp}
\end{figure}

In Figs.~\ref{fig:fig-ness-th} and~\ref{fig:fig-ness-temp}, we demonstrate an excellent agreement between theory and simulations for given values of $(m,T,\sigma)$. From the figure, it is evident that our
analytical approach works very well for both small and large values of $m$.

The ratio $p(\theta)/n(\theta)$ gives the temperature $T(\theta)$.
Equilibrium state of a system necessarily implies a spatially uniform
temperature profile, i.e., $T(\theta)$ equals the temperature $T$,
independent of $\theta$, where $T$ is the temperature of the
heat bath the system is in contact with. The spatially non-uniform
temperature profile in the right panel of Fig.~\ref{fig:fig-ness-temp}
lends further credence to the suggestion that the synchronized state we are
dealing with is a NESS. The figure also shows a density-temperature anticorrelation, i.e.,
the temperature is peaked at a value of $\theta$ at which the density
is minimum, and vice versa. This phenomenon of temperature inversion has
been argued to be a generic feature of long-range interacting systems in
NESSs~\cite{Casetti:2014,Teles:2015,Gupta:2016}.

\section{Conclusions}
\label{sec:conclusions}
In this review, we presented an overview of statistical mechanical
aspects of large networks of coupled phase oscillators with distributed
natural frequencies. We
analyzed an issue of both theoretical and practical relevance, namely,
the conditions under which the system displays the emergent phenomenon
of spontaneous synchronization, whereby a macroscopic population of
oscillators exhibits in-phase oscillations. Considering a general
unimodal distribution of the natural frequencies, we discussed about
phase transitions that occur between a synchronized
phase and an unsynchronized/incoherent phase on tuning of dynamical
parameters. While the initial part
of the review focussed on the celebrated Kuramoto model involving
first-order overdamped dynamics of a system of globally-coupled phase
oscillators, the central part was devoted to discussing recent results
obtained for a generalized Kuramoto model that includes effects of
inertial terms and stochastic noise, with the underlying dynamics being
second order in time.  In the limit of zero
noise and inertia, the dynamics reduces to that of the Kuramoto
model, while at finite noise and inertia but in the absence of natural frequencies, the dynamics becomes
the canonical ensemble dynamics of a paradigmatic model to study static and dynamic properties of
long-range interacting systems, namely, the Hamiltonian mean-field (HMF) model.
For the generalized model, we discussed how a
combination of competing dynamical effects results in a rather rich and
complex phase diagram in the stationary state. In particular, for a
general unimodal frequency distribution, we reported the complete phase diagram of the
model, and demonstrated that the system undergoes a nonequilibrium first-order phase transition from a synchronized
phase at low values of the dynamical parameters to an incoherent phase
at high values. In proper limits, the phase diagram incorporates the
known phase transitions of the Kuramoto and the HMF model. Following the
work on the generalized model reported in this review, there has been a huge surge
in interest in studying the model and its extension, leading to a number
of recent publications in the area. Some representative ones are
Refs.~\cite{Komarov:2014,Olmi:2014,Olmi:2015,Olmi:2015-1,Jorg:2015,Barre:2016,Chen:2017,Yuan:2017}.
This review was entirely devoted to studies of mean-field interaction
between the oscillators, namely, the case where every oscillator
interacts with every other with a strength that is the same for every
pair, thereby representing an extreme case of long-range interactions.
However, to model specific situations of interest, the setup has also been
generalized to consider the case in which the oscillators interact with one
another with a strength that decays with the spatial separation between
the oscillators~\cite{Rogers:1996}. Recent results within such a setup
and with a focus similar to the present review may be found in
Refs.~\cite{Gupta:2012a,Gupta:2012b}.

In conclusion, we believe that a statistical mechanical approach to
study a system of globally-coupled phase oscillators provides a useful tool for
investigating the collective behavior of the system, and allows to
deepen our understanding of peculiar features of nonequilibrium
stationary states {\it vis-\`{a}-vis} equilibrium, besides offering new
and exciting opportunities of experimental exploration.

\section*{Acknowledgements}
Stefano G is grateful to Giacomo Innocenti for useful discussions on the
Kuramoto model. Shamik G especially thanks Alessandro Campa for several useful
and insightful discussions and comments on the Kuramoto model, and, in
particular, on its derivation using the phase approximation technique as discussed
in this review. We thank the Max Planck Institute for the Physics of
Complex Systems, Dresden, Germany, for the hospitality during the
workshop ``Dynamics of Coupled Oscillators: 40 years of the Kuramoto
model,'' where this paper was conceptualized.


\end{document}